\documentclass[10pt]{iopart}
\usepackage{iopams}
\usepackage{times}
\usepackage{colordvi}

\def\de{\partial} 

\begin{document}

\title[]{Gauge-invariant Non-spherical Metric
Perturbations of Schwarzschild Black-Hole Spacetimes}

\author{Alessandro Nagar\dag$^*$\ and Luciano Rezzolla\ddag\S }

\address{\dag Department of Astronomy and Astrophysics, University of Valencia,
Burjassot, Spain}

\address{$^*$Department of Physics at the Politecnico, Turin, Italy}

\address{\ddag SISSA, International School for Advanced Studies and INFN, 
Trieste, Italy}

\address{\S Department of Physics, Louisiana State University, Baton
        Rouge, USA}

\begin{abstract}
  	The theory of gauge-invariant non-spherical metric perturbations
	of Schwarzschild black hole spacetimes is now well established. 
	Yet, as different notations and conventions have been used throughout 
	the years, the literature on  the subject is  often confusing and
	sometimes confused.  The  purpose of this paper is  to review and
	collect the relevant expressions related to the Regge-Wheeler and
	Zerilli equations for the  odd and even-parity perturbations of a
	Schwarzschild spacetime.   Special attention is paid  to the form
	they assume  in the presence  of matter-sources and, for  the two
	most  popular conventions  in the  literature, to  the asymptotic
	expressions and  gravitational-wave amplitudes.  Besides pointing
	out  some  inconsistencies  in  the literature,  the  expressions
	collected  here  could  serve   as  a  quick  reference  for  the
	calculation  of the  perturbations of Schwarzschild black hole 
	spacetimes driven  by generic  sources  and for  those approaches 
	in  which gravitational waves  are   extracted  from  numerically  
	generated spacetimes.
\end{abstract}





\section{Introduction}

	Black-hole perturbation theory (see for example the monograph by
Chandrasekhar~\cite{chandra}, the book by Frolov and
Novikov~\cite{frolov98} or the recent review by Kokkotas and
Schmidt~\cite{ks99}) has been fundamental not only for
understanding the stability and oscillations properties of black hole
spacetimes~\cite{rw57}, but also as an essential tool for clarifying the
dynamics that accompanies the process of black hole formation as a result
of gravitational collapse~\cite{price1972a, price1972b}. As one example
among the many possible, the use of perturbation theory has led to the
discovery that Schwarzschild black holes are characterised by decaying
modes of oscillation that depend on the black hole mass only, i.e.
the black hole quasi-normal modes~\cite{vish70a, vish70b,
press71,chandra75}.  Similarly, black-hole perturbation theory and the
identification of a power-law decay in the late-time dynamics of generic
black-hole perturbations has led to important theorems, such as the `no
hair' theorem, underlining the basic black-hole property of removing all
perturbations so that `all that can be radiated away is radiated
away'~\cite{price1972a, price1972b, mtw74}.

	The foundations of non-spherical metric perturbations of
Schwarzschild black holes date back to the work in 1957 of Regge and
Wheeler~\cite{rw57}, where the linear stability of the Schwarzschild
singularity was first addressed. A number of investigations, both
gauge-invariant and not, then followed in the 1970s, when many different
approaches were proposed and some of the most important results about the
physics of perturbed spherical and rotating black holes
established~\cite{price1972a, price1972b,vish70a, vish70b,
chandra75,zerilli70c, zerilli70b, moncrief74, CPM78, CPM79, teukolsky72,
teukolsky73}. Building on these studies, which defined most of the
mathematical apparatus behind generic perturbations of black holes, a
number of applications have been performed to study, for instance, the
evolutions of perturbations on a collapsing background
spacetime~\cite{gerlach79a, gerlach79b, gerlach80, karlovini02, seidel87,
seidel88, seidel90, seidel91}. Furthermore, the gauge-invariant and
coordinate independent formalism for perturbations of spherically
symmetric spectimes developed in the 70's by Gerlach and Sengupta in
\cite{gerlach79a, gerlach79b, gerlach80}, has been recently extended to
higher-dimensional spacetimes with a maximally symmetric subspace 
in~\cite{kodama00, kodama03a, kodama03b, kodama04}, for the study of
perturbations in brane-world models.

	Also nowadays, when numerical relativity calculations allow to
evolve the Einstein equations in the absence of symmetries and in fully
nonlinear regimes, black hole perturbative techniques represent important
tools. Schwarzschild perturbation theory, for instance, has been useful
in studying the late-time behaviour of the coalescence of compact
binaries in a numerical simulation after the apparent horizon has
formed~\cite{pp94, ac94, ast95}. In addition, methods have been developed
that match a fully numerical and three-dimensional Cauchy solution of
Einstein's equations on spacelike hypersurfaces with a perturbative
solution in a region where the components of three-metric (or of the
extrinsic curvature) can be treated as linear perturbations of a
Schwarzschild black hole (this is usually referred to as the
`Cauchy-Perturbative matching')~\cite{bbhgca1,retal98, cs99, acs98,
Rezzolla99a}. This method, in turn, allows to ``extract'' the
gravitational waves generated by the simulation, evolve them out to the
wave-zone where they assume their asymptotic form, and ultimately provide
outer boundary conditions for the numerical evolution.

	Overall, therefore, black hole perturbation theory represents a
very powerful tool with multiple applications and developments. The
purpose of this review is to review the theory of metric perturbations of
Schwarzschild black holes, especially in its gauge-invariant
formulations. While a systematic derivation of the most relevant
expressions is provided, special care is paid to `filter' those
technical details that may obscure the important results and provide the
reader with a set of expressions that can be readily used for the
calculation of the odd and even-parity perturbations of a Schwarzschild
spacetime in the presence of generic matter-sources. Also, an effort is
made to `steer' the reader through the numerous conventions and
notations that have accompanied the development of the formalism over the
years, pointing out the misprints and errors that can be found in the
literature. Finally, although most of what presented here is not new,
some expressions that have not been discussed before in the literature
are also introduced.

	It should be noted that our discussion will be restricted
exclusively to black-hole spacetimes, although a Schwarzschild metric can
also be used to describe the vacuum exterior of a spherical relativistic
star. The reason behind this choice is that boundary conditions different
from the ones used for black holes need to be imposed at the stellar
surface. Furthermore, in the case of a relativistic star the Einstein
equations need to be suitably coupled with the equations of relativistic
hydrodynamics. A detailed discussion of how to do this in practice is
beyond the purpose of this review, which is instead focused on a vacuum
background spacetime. However, a recent review on the theory of
relativistic stellar perturbations can be found in~\cite{ks99}.

	The organisation of the paper is as follows. In
section~\ref{gi_intro} we review the basic properties and advantages of a
gauge-invariant description of the perturbations of a curved spacetime,
while in section~\ref{sec:multipole} we review the multipolar expansion
of the perturbations of a Schwarzschild spacetime. Sections~\ref{opp}
and~\ref{epp} are instead devoted to the derivation of the general
equations governing the evolution of odd and even-parity perturbations,
respectively, whose asymptotic expressions for the gravitational-wave
amplitudes and losses are presented in section~\ref{ae}. Finally,
section~\ref{conclusions} contains our conclusions and a number of
Appendices follow to provide some useful relations and expressions for
the angular parts of the perturbations.

	We use a spacelike signature $(-,+,+,+)$ with Greek indices
running from 0 to 3 and indicate partial and covariant derivatives in the
$\mu$-th direction with the symbols $\partial_{\mu}$ and
$\nabla_{\!\mu}$, respectively. Also, symmetrized and antisymmetrized 
indices in a tensor will be indicated with round and square brackets,
respectively.

\section{Gauge-invariant metric perturbations}
\label{gi_intro}

	To highlight the importance of gauge-invariant quantities in
describing the metric perturbations it is useful to recall that even if
the coordinate system of the background spacetime has been fixed, the
coordinate freedom of general relativity introduces a problem when linear
perturbations are added. In particular, it is not possible to distinguish
an infinitesimal `physical' perturbation from one produced as a result
of an infinitesimal coordinate transformation (or
gauge-transformation). This difficulty, however, can be removed either 
by explicitly fixing a gauge (as done in the pioneering works on the 
subject~\cite{rw57, price1972a, price1972b, vish70a, vish70b, zerilli70c, 
zerilli70b}), or by introducing linearly gauge--invariant perturbations 
(as initially suggested by Moncrief~\cite{moncrief74} and subsequently 
adopted in several applications~\cite{CPM78, CPM79, seidel87, seidel88, 
seidel90,seidel91}).

	More specifically, given a tensor field ${\boldsymbol X}$ in a
background metric ${\boldsymbol g{\!\!\!\!\;
\raisebox{-0.1ex}{$^{^0}$}}}$ and $\delta {\boldsymbol X}$ its
infinitesimal perturbation, an infinitesimal coordinate transformation
$x^{\mu}\rightarrow x^{\mu'} \equiv x^{\mu}+\xi^{\mu}$ with $\xi^{\mu}\ll
1$ will yield a new tensor field
\begin{equation}
\delta {\boldsymbol X}\rightarrow \delta {\boldsymbol X}'=\delta
	{\boldsymbol X}+ {\cal L}_{\boldsymbol \xi} {\boldsymbol X}\ ,
\end{equation}
where ${\cal L}_{\boldsymbol \xi}$ is the Lie derivative along
${\boldsymbol \xi}$ in the metric ${\boldsymbol g{\!\!\!\!\;
\raisebox{-0.1ex}{$^{^0}$}}}$. We will then consider $\delta {\boldsymbol
X}$ to be gauge-invariant if and only if ${\cal L}_{\boldsymbol
\xi}{\boldsymbol X}=0$.

	Stated differently, the possibility of building gauge--invariant
metric perturbations relies on the existence of symmetries of the
background metric. In the case of a general spherically symmetric
background spacetime (i.e. one allowing for a time dependence) and
which has been decomposed in multipoles (see Sect.~\ref{sec:multipole}),
the construction of gauge-invariant quantities is possible for multipoles
of order $\ell \geq 2$ only~\cite{gerlach79a, gerlach79b, martingarcia99,
gundlach00}.  In practice, the advantage in the use of gauge-invariant
quantities is that they are naturally related to scalar observables and,
for what is relevant here, to the the energy and momentum of
gravitational waves. At the same time, this choice guarantees that
possible gauge-dependent contributions are excluded by construction.

	Of course, this procedure is possible if and only if the
background metric has the proper symmetries under infinitesimal
coordinates transformation and a gauge-invariant formulation of the
Einstein equations for the perturbations of a general spacetime is not
possible. Nevertheless, since any asymptotically flat spacetime can in
general be matched to a Schwarzschild one at sufficiently large
distances, a gauge-invariant formulation can be an effective tool to
extract physically information about the gravitational waves generated in
a numerically evolved, asymptotically flat spacetime~\cite{bbhgca1,
retal98, cs99, acs98, Rezzolla99a}.

\section{Multipolar expansion of metric perturbations}
\label{sec:multipole}

	Given a spherically symmetric Schwarzschild solution with metric
${\boldsymbol g{\!\!\!\!\; \raisebox{-0.1ex}{$^{^0}$}}}$ and line element
\begin{equation}
ds^2 \equiv g{\!\!\!\!\; \raisebox{-0.1ex}{$^{^0}$}}_{\mu\nu}
	dx^{\mu}dx^{\nu}
	=-e^{2a}dt^2+e^{2b}dr^2+r^2\left(d\theta^2+\sin^2\theta
	d\phi^2\right)\ , 
\end{equation}
where $e^{2a}=e^{-2b}=\left(1-2M/r\right)$, we generically consider small
non-spherical perturbations $h_{\mu \nu}$ such that the new perturbed
metric is
\begin{equation}
\label{gmunu}
g_{\mu\nu} \equiv  g{\!\!\!\!\; \raisebox{-0.1ex}{$^{^0}$}}_{\mu\nu} + 
	h_{\mu\nu}\ ,
\end{equation}
where $|h_{\mu \nu}|/|g{\!\!\!\!\; \raisebox{-0.1ex}{$^{^0}$}}_{\mu\nu}|
\ll 1$. It should be noted that although we have here chosen to employ
Schwarzschild coordinates to facilitate the comparison with much of the
previous literature, this is not the only possible choice. Indeed, it is
possible to formulate the perturbations equations independently of the
choice of coordinates as discussed in~\cite{mp_05} and in~\cite{sarbach01}.

	Because the background manifold ${\cal M}$ is spatially
spherically symmetric, it can be written as the product ${\cal M}={\sf
M}^2\times {\sf S}^2$, where ${\sf M}^2$ is a Lorentzian 2-dimensional
manifold of coordinates $(t,r)$ and ${\sf S}^2$ is the 2-sphere of unit
radius and coordinates $(\theta, \phi)$. As a result of this
decomposition, the perturbations can be split {\it ab initio} in a
part confined to ${\sf M}^2$ and in a part confined on the 2-sphere ${\sf
S}^2$ of metric ${\boldsymbol \gamma}$. Exploiting this, we can expand
the metric perturbations ${\boldsymbol h}$ in multipoles referred to as
odd or even-parity according to their transformation properties under
parity. In particular, are {\it odd} (or {\it axial}) multipoles those
that transform as $(-1)^{\ell+1}$, under a parity transformation
$(\theta, \phi) \to (\pi-\theta, \pi +\phi)$, while are {\it even} (or
{\it polar}) those multipoles that transform as $(-1)^{\ell}$. As a
result, the metric perturbations can be written as
\begin{equation}
\label{hmunu}
h_{\mu\nu}= 
	\sum_{\ell,m}\left[\left(h_{\mu\nu}^{{\ell m}}\right)^{({\rm o})} 
	+\left(h_{\mu\nu}^{\ell m}\right)^{({\rm e})}\right]\ ,
\end{equation}
where
$$
\sum_{\ell,m} \equiv \sum_{\ell=2}^{\infty} \;
\sum_{m=-\ell}^{\ell} \nonumber \ . 
$$
Introducing now a notation inspired by that of
Gerlach and Sengupta~\cite{gerlach79a, gerlach79b, gerlach80} and
recently revived by Gundlach and Mart\'in-Garc\'ia~\cite{martingarcia99,
gundlach00, gundlach01}, we use upper-case indices $A,\,B\, \ldots=0,1$
to label the coordinates of ${\sf M}^2$ and lower-case indices
$c,d\ldots=2,3$ to label the coordinates of ${\sf S}^2$.

	Using this notation, the scalar spherical harmonics $Y^{{\ell
m}}$ are then simply defined as
\begin{equation}
\label{def:harmonics}
\gamma^{cd}\nabla_d\nabla_cY^{{\ell m}}= -\Lambda Y^{{\ell m}}\ ,
\end{equation}
where $\nabla_c$ indicates the covariant derivative with respect to the
metric ${\boldsymbol \gamma}\equiv \mathrm{diag}(1,\sin^2\theta)$ of
${\sf S}^2$ and where $\Lambda \equiv \ell(\ell+1)$. It is now
convenient to express the odd and even-parity metric functions in
(\ref{gmunu}) in terms of tensor spherical harmonics. To do this we
introduce the axial vector $S^{{\ell m}}_c$ defined as
\begin{equation}
S_c^{{\ell m}}
\equiv \epsilon_{cd}\gamma^{de}\nabla_e Y^{{\ell m}}\ , 
\end{equation}
where $\epsilon_{cd}$ is the volume form on ${\sf S}^2$ as defined by the
condition $\epsilon_{cd}\epsilon^{ce}=\gamma_d^{\;\;e}$ and such that
$\nabla_c\epsilon_{ab}=0$. In this way, each odd-parity metric function
in (\ref{gmunu}) can be written as
\begin{eqnarray}
\label{odd:metric}
\left(h_{\mu\nu}^{{\ell m}}\right)^{({\rm o})}= \left(\begin{array}{cc} 0 &
	h_A^{({\rm o})}S_{c}^{{\ell m}} \\ 
	\cr h_A^{({\rm o})}S_{c}^{{\ell m}} & h
	\nabla_{(d} S_{c)}^{{\ell m}}\\
\end{array}
\right)\ ,\\ \nonumber
\end{eqnarray}
where $h_{A}^{({\rm o})},\, h$ are functions of $(t,r)$ only and where we
have omitted the indices $\ell, m$ for clarity.

	Proceeding in a similar manner, each even-parity metric function
can be decomposed in tensor spherical harmonics as
\begin{eqnarray}
\label{even:metric}
\fl \qquad \left(h_{\mu\nu}^{{\ell m}}\right)^{(\mathrm{e})}
	=\left(\begin{array}{cc|cc} e^{2a} H_0Y^{{\ell m}} & 
	H_1Y^{{\ell m}} & \qquad
	h_A^{({\rm e})}\nabla_{c}Y^{{\ell m}} \\ 
	H_1Y^{{\ell m}} & e^{2b} H_2Y^{{\ell m}} & \\ 
\hline & &\\ 
	h_A^{({\rm e})}\nabla_c Y^{{\ell m}} & &
	r^2\left(KY^{{\ell m}}\gamma_{cd}+
	G \nabla_d\nabla_c Y^{{\ell m}}\right)\\
\end{array}\right)\ ,\\
\nonumber
\end{eqnarray}
where $H_0,\,H_1,\,H_2,\, h_0^{({\rm e})},\,h_1^{({\rm e})}\, K,\,G$
(with the indices $\ell, m$ omitted for clarity) are the coefficients of
the even-parity expansion, are also functions of $(t,r)$ only.

	Note that we have here used the Regge-Wheeler set of tensor
harmonics to decompose the even-parity part of the metric in
multipoles~\cite{rw57}. Despite this being a popular choice in the
literature, it is not the most convenient one since the tensor harmonics
in this set are not linearly independent. An orthonormal set is instead
given by the Zerilli-Mathews tensor harmonics \cite{zerilli70a,mathews62}
and the transformation from one basis to the other is given by defining
the tensor $Z_{cd}^{{\ell m}}$ confined on the 2-sphere ${\sf S}^2$
\begin{equation}
Z_{cd}^{{\ell m}} \equiv \nabla_c \nabla_d Y^{{\ell m}}+
	\frac{\Lambda}{2}\gamma_{cd}Y^{{\ell m}}\ ,
\end{equation}
and then replacing in equation~(\ref{even:metric}) the second covariant
derivative of the spherical harmonics $\nabla_c\nabla_dY^{{\ell m}}$ with
$Z^{{\ell m}}_{cd}$. This transformation has to be taken into account,
for instance, when developing gauge-invariant procedures for extracting
the gravitational-wave content from numerically generated spacetimes
which are ``almost'' Schwarzschild spacetime~\cite{abrahams92, anninos93,
anninos95a, anninos95b, abrahams96}.

	Besides vacuum tensor perturbations, the background Schwarzschild
spacetime can be modified if non-vacuum tensor perturbations are present
and have a mass-energy much smaller than that of the black hole but
sufficiently large to perturb it. In this case, the generic stress-energy
tensor $t_{\mu\nu}$ describing the matter-sources can be 
similarly decomposed in odd and even-parity parts 
\begin{equation}
t_{\mu\nu}=\sum_{\ell,m}\left[\left(t^{{\ell m}}_{\mu\nu}\right)^{({\rm
	o})}+\left(t^{{\ell m}}_{\mu\nu}\right)^{({\rm e})}\right]
	\ ,
\end{equation}
that are naturally gauge-invariant since the background is the vacuum
Schwarzschild spacetime and are given explicitely by
\begin{eqnarray}
\label{source:odd}
\left(t^{{\ell m}}_{\mu\nu}\right)^{({\rm o})}= \left(
\begin{array}{cc}
	0 & L_A^{{\ell m}}S_{c}^{{\ell m}} \\ \\ 
	L_A^{{\ell m}}S_{c}^{{\ell m}} &
	L^{{\ell m}} \nabla_{(d}S_{c)}^{{\ell m}}\\
\end{array}
\right)\ ,\\ \nonumber
\end{eqnarray}
for the odd-parity part and by
\begin{eqnarray}
\label{source:even}
\left(t^{{\ell m}}_{\mu\nu}\right)^{({\rm e})}= 
	\left(\begin{array}{c|c}
	T_{AB}^{{\ell m}}Y^{{\ell m}} & 
	T_A^{{\ell m}}\nabla_c Y^{{\ell m}}\\ 
	& \\ \hline & \\ 
	T_A^{{\ell m}}\nabla_c
	Y^{{\ell m}} & r^2T_3^{{\ell m}}Y^{{\ell m}}\gamma_{cd}+
	T_2^{{\ell m}}Z_{cd}^{{\ell m}}\\
\end{array}
\right) \ , \\ \nonumber
\end{eqnarray}
for the even-parity one. Note that we have now used the Zerilli-Matthews
set of harmonics for the expansion, that the ten coefficients
$L_{A}^{{\ell m}},\, L^{{\ell m}},\, T_{AB}^{{\ell m}},\, T_{A}^{{\ell
m}},\, T_2^{{\ell m}},\, T_3^{{\ell m}}$ are gauge-invariant and that
explicit expressions for them will be presented in the following Sections.

        With the perturbations decomposed in a convenient form, we can
now consider the Einstein field equations that, in the static vacuum
background, take the simple form
\begin{equation}
\label{efe_u}
R{\!\!\!\!\; \raisebox{0.3ex}{$^{^0}$}}_{\mu \nu} = 0 \ ,
\end{equation}
where ${\boldsymbol R}{\!\!\!\!\!\; \raisebox{0.3ex}{$^{^0}$}}$ is the
Ricci tensor built from the background metric ${\boldsymbol g}{\!\!\!\!\;
\raisebox{-0.1ex}{$^{^0}$}}$. At first order in the perturbations, the
linearity can be exploited to break up the Einstein equations as
\begin{equation}
\label{efe_vp}
R{\!\!\!\!\; \raisebox{0.3ex}{$^{^0}$}}_{\mu \nu}+R_{\mu\nu}-
\frac{1}{2}g{\!\!\!\!\; \raisebox{-0.1ex}{$^{^0}$}}_{\mu\nu}R=8\pi t_{\mu\nu} \ ,
\end{equation}
where ${\boldsymbol R}$ is now the Ricci tensor built from the metric
perturbations ${\boldsymbol h}$. Using equations (\ref{efe_u}), the field
equations then reduce to
\begin{equation}
\label{efe_p}
R_{\mu\nu}-\frac{1}{2}g{\!\!\!\!\; \raisebox{-0.1ex}{$^{^0}$}}_{\mu\nu}R
=8\pi t_{\mu\nu} \ ,
\end{equation}
and where the components of the Ricci tensor are explicitely given by
\begin{equation}
\label{deltarmunu}
R_{\mu \nu}=   \nabla_{\beta} \Gamma_{\mu \nu}^{\beta} -\nabla_{\nu} \Gamma_{\mu \beta}^{\beta} \ .
\end{equation}
Here, ${\boldsymbol \Gamma}$ are the Christoffel symbols relative to the
perturbed metric ${\boldsymbol h}$, i.e.
\begin{equation}
\Gamma^{\beta}_{\ \ \mu \nu} = \frac{1}{2} 
	g{\!\!\!\!\; \raisebox{-0.1ex}{$^{^0}$}
	\raisebox{-0.1ex}{$^{\alpha\beta}$}}
        \left( \partial_{\nu} h_{\mu \alpha} +
        \partial_{\mu}        h_{\nu \alpha} - 
	\partial_{\alpha}     h_{\mu \nu}       \right) \ ,
\end{equation}
and should be distinguished from the corresponding Christoffel symbols
${\boldsymbol \Gamma}{\!\!\!\!\!\; \raisebox{0.3ex}{$^{^0}$}}$ relative to
the background metric ${\boldsymbol g}{\!\!\!\!\;
\raisebox{-0.1ex}{$^{^0}$}}$.

	Note that while a generic perturbation will be a mixture of odd
and even-parity contributions, we will exploit the linearity of the
approach to handle them separately and simplify the treatment.  In the
following two Sections we will discuss the form the Einstein equations
(\ref{efe_p}) assume in response to odd and even-parity perturbations
over a Schwarzschild background. In particular, we will show how the
three odd-parity coefficients of the expansion in harmonics of the metric
and the seven even-parity ones can be combined to give two
gauge-invariant master equations, named after Regge and
Wheeler~\cite{rw57} and Zerilli~\cite{zerilli70a}, each of which is a
wave-like equation in a scattering potential\footnote{These results were
originally obtained by Regge and Wheeler~\cite{rw57} and by
Zerilli~\cite{zerilli70c, zerilli70b} in a specific gauge (i.e. the
Regge-Wheeler gauge). Subsequently, the work of Moncrief showed how to
reformulate the problem in a gauge-invariant form by deriving the
equations from a suitable variational principle~\cite{moncrief74}. Some
details of Moncrief's approach to odd-parity perturbations are briefly
described in~\ref{appendix_d}.}.

 	Although our attention is here focussed on the $radiative$
degrees of freedom of the perturbations (i.e. those with $\ell\ge
2$) because of their obvious application to the modelling of sources of
gravitational waves, a comment should be made also on lower-order
multipoles. In particular, it is worth remarking that the monopole
component of the metric for a vacuum perturbation (i.e.  with $\ell
= 0$) is only of even-parity type and represents a variation in the
mass-parameter of the Schwarzschild solution. On the other hand, the
dipole component of the even-parity metric for a vacuum perturbation
(i.e. with $\ell=1$) is of pure-gauge type and it can be removed by
means of a suitable gauge transformation~\cite{zerilli70b}. Note that
this is not the case for a dipolar odd-parity metric perturbation, which
can instead be associated to the introduction of angular momentum onto
the background metric. A detailed discussion of these low multipoles can
be found in Appendix G of~\cite{zerilli70b} and in~\cite{mp_05}.

\section{Gauge-invariant odd-parity perturbations}
\label{opp}

	Before discussing the derivation of the odd-parity equation a
choice should be made for the odd-parity master function. Unfortunately,
this choice has not been unique over the years and different authors have
made different choices, making comparisons between different approaches
less straightforward. Here, we will make a choice which highlights the
relation with the gravitational-wave amplitude measured by a distant
observer and, in particular, we construct the gauge-invariant combination
of multipoles~\cite{gerlach79a, martingarcia99, harada03}
\begin{equation}
\label{def:kA}
k_{A} \equiv h_A-\nabla_{\!A} h +2h\frac{\nabla_{\!A}r}{r}\ ,
\end{equation}
where, we recall, $\nabla_{\!A}$ represents the covariant derivative
with respect to the connection of the submanifold ${\sf M}^2$. In this
way, the function
\begin{equation}
\label{phi:GM}
\Phi^{({\rm o})}(t,r) \equiv
	r^3\epsilon^{AB}\nabla_{\!B}\left(\frac{k_A}{r^2}\right)
	= r\left[\partial_{t}h_{1}^{({\rm o})}-
	r^2\partial_r\left(\frac{h_0^{({\rm o})}}{r^2}\right)\right]\ ,
\end{equation}
where $\epsilon_{AB}$ is the antisymmetric volume form on ${\sf M}^2$
($\epsilon^{01} = - \epsilon^{10} =- 1$, see~\ref{appendix_d}), 
is gauge-invariant and will be our choice for the Regge-Wheeler master 
function~\cite{gerlach79a,martingarcia99, gundlach00, harada03}.

	A slight variation of the master function (\ref{phi:GM}) has been
introduced by Cunningham, Price and Moncrief~\cite{CPM78} in terms of the
function $\widetilde{\psi} \equiv \Lambda \Phi^{({\rm o})}$ and this has
been used so extensively in the literature~\cite{seidel87, seidel88,
seidel91} that it is now commonly referred to as the
Cunningham-Price-Moncrief (CPM) convention.  We partly follow this
suggestion and introduce a new master function for the odd-parity
perturbations defined as
\begin{equation}
\label{CPM_n}
\Psi^{({\rm o})} \equiv \frac{1}{\Lambda-2} \,\Phi^{({\rm o})}\ .
\end{equation}

	With the choice (\ref{CPM_n}), the Einstein field equations
(\ref{efe_p}) with odd-parity perturbations lead to the inhomogeneous
`Regge-Wheeler' equation
\begin{equation}
\label{rw_eq}
\partial^2_{t} \Psi^{({\rm o})} - \partial^2_{r_*} \Psi^{({\rm o})} + 	
	V_{\ell}^{({\rm o})}\Psi^{({\rm o})}=S^{({\rm o})}\ ,
\end{equation}
where 
\begin{equation}
r_*\equiv r+2M\ln \left( \frac{r}{2M}-1\right) \ ,
\end{equation}
is the `tortoise coordinate'~\cite{mtw74} and $V_{\ell}^{({\rm o})}$ is
the odd-parity potential, defined as
\begin{equation}
\label{odd:potential}
V_{\ell}^{({\rm o})} \equiv \left(1-\frac{2M}{r}\right)
	\left(\frac{\Lambda}{r^2}- \frac{6M}{r^3} \right)\ .
\end{equation}
The right-hand-side of equation~(\ref{rw_eq}) represents the generic
odd-parity ``matter-source'' and is given by
\begin{equation}
\label{odd:source}
\fl S^{({\rm o})} \equiv 
	\frac{16\pi r}{\Lambda-2} \, e^{2a}\epsilon^{AB}\nabla_{\!B}L_{A} = 
	\frac{16\pi r}{\Lambda-2} \left[\left(1-\frac{2M}{r}\right)
	\partial_{t}L_{1}^{{\ell m}} - \partial_{r_*}
	L_{0}^{{\ell m}}\right]\ , 
\end{equation}
with the components of the odd-parity matter-source vector defined as
\begin{equation}
L_{A}^{{\ell m}} \equiv \frac{1}{\Lambda}\int
	\frac{d\Omega}{\sin\theta}\left(\mathrm{i}\,m\,
	t_{A2}Y^{*}_{{\ell m}}+
	t_{A3}\,\partial_{\theta}Y_{\ell m}^{*}\right)\ , \qquad A=0,1\ ,
\end{equation}
and where $d\Omega=\sin\theta d\phi d\theta$ is the surface element on
the 2-sphere ${\sf S}^2$. 

	As a result of this construction, given a generic odd-parity
matter-source $t_{\mu \nu}$ whose dynamics is known ({\it e.g.}  through
the solution of the equations of motion if the source is point-like, or
those of relativistic hydrodynamics if the source is extended), equation
(\ref{rw_eq}) describes the evolution of the perturbations induced on the
black-hole spacetime by the dynamics of the matter-source.

	It should be noted that another choice for the gauge-invariant
odd-parity master variable is possible and indeed was originally proposed
by Moncrief~\cite{moncrief74}. This function, which hereafter we will
refer to as the odd-parity Moncrief function, is defined as
\begin{equation}
\label{mfop}
Q^{(\rm o)}\equiv \frac{\nabla_{\!A} r}{r}
	g{\!\!\!\!\; \raisebox{-0.1ex}{$^{^0}$}}^{AB}k_B =
	\frac{1}{r}\left(1-\frac{2M}{r}\right)
	\left[h_{1}^{({\rm o})}+\frac{r^2}{2} \partial_r
	\left(\frac{h_2}{r^2}\right)\right]\ ,
\end{equation}
where the first expression is coordinate independent~\cite{mp_05}, while
the second one is specialized to Schwarzschild coordinates with $h_2 = -
2 h$~\cite{moncrief74}. In the Regge-Wheeler gauge $h_2=h=0$, and the
definition (\ref{mfop}) coincides with the variable used by Regge and
Wheeler~\cite{rw57}. Historically, the choice of (\ref{mfop}) as master
variable has been the most common in the literature to describe
odd-parity perturbations of a Schwarzschild spacetime and we will refer
to it as `Regge-Wheeler' (RW) convention. It should be noted that while
(\ref{mfop}) is a solution of the Regge-Wheeler equation, the
corresponding source term differs from expression (\ref{odd:source}). A
general expression of the source in the RW convention can be found 
in~\cite{mp_05,martel04} together with its specification for a
point-particle (see also~\cite{ap99, tominaga99, ferrari00}).

	The two master functions $Q^{(\rm o)}$ and $\Psi^{(\rm o)}$ are
intimately related in at least two different ways. Firstly, through the
variational formalism employed by Moncrief in~\cite{moncrief74} and
a brief discussion of these relations is presented
in~\ref{appendix_d}. Secondly, through an explicit expression that
reads~\cite{mp_05}
\begin{equation}
\label{Q_vs_Psi}
	\de_t\Psi^{(\rm o)}=-Q^{(\rm o)} +
	\frac{16\pi}{\Lambda-2}\frac{r}{e^{2b}} L_1^{\ell m} \ .
\end{equation} 
Equation (\ref{Q_vs_Psi}) highlights an important difference between the
two master functions which is not just a dimensional one 
(i.e. $\Psi^{({\rm o})}$ has the dimensions of a length, while $Q^{({\rm o})}$
is dimensionless) and that will have consequences on the asymptotic
expressions for the gravitational waveforms when these are expressed in
one or in the other convention. A detailed discussion of this will be
made in sections.~\ref{sec:h_from_op}, \ref{a_general} and
\ref{e_am_losses}.

\section{Gauge-invariant even-parity perturbations}
\label{epp}

	Also in the case of even-parity perturbations, an evolution
equation similar to the Regge-Wheeler one (\ref{rw_eq}) can be found. In
particular, following Moncrief \cite{moncrief74}, we define the functions
\begin{eqnarray}
\label{kappa1}
\kappa_1 & \equiv & K+\frac{1}{e^{2b}}\left(r\partial_r G-
	\frac{2}{r}h_1^{({\rm e})}\right)\ ,\\
\label{kappa2}
\kappa_2 & \equiv &\frac{1}{2}\left[e^{2b}H_2-
	e^b \partial_r \left(r e^b K\right)\right]\ ,
\label{def:q1}
\end{eqnarray}
where both $\kappa_1$ and $\kappa_2$ are gauge-invariant functions, as
well as the following linear combination
\begin{equation}
\label{q1}
q_1  \equiv r\Lambda\kappa_1 + \frac{4r}{e^{4b}}\kappa_2 \ .
\end{equation}
Strictly related to expression (\ref{q1}) is the gauge-invariant function
most frequently used in the literature~\cite{ seidel90, gundlach01,
martel04,lousto97, ruoff01a, ruoff01b, martel02, nagar04,
poisson04}\footnote{Note that expression (\ref{psi_e}) corrects 
the sign in the corresponding definition of the even-parity function 
used in~\cite{ko04} (cf~equation(7) in~\cite{ko04}).}
\begin{equation}
\label{psi_e}
{\Psi}^{({\rm e})} \equiv \frac{r
	q_1}{\Lambda\left[r\left(\Lambda-2\right)+6M\right]} \ .
\end{equation}
which is also the solution of the inhomogeneous even-parity master
equation or `Zerilli' equation
\begin{equation}
\label{zerilli_eq}
\partial^2_{t}{\Psi}^{({\rm e})} - 
	\partial^2_{r_*}{\Psi}^{({\rm e})}+
	V_{\ell}^{({\rm e})}{\Psi}^{({\rm e})}=S^{({\rm e})} \ ,
\end{equation}
and, again, is a wave-like equation in the scattering Zerilli
potential~\cite{zerilli70c}
\begin{equation}
\fl V_{\ell}^{({\rm e})} \equiv
	\left(1-\frac{2M}{r}\right)\frac{\Lambda(\Lambda-2)^2r^3
	+6(\Lambda-2)^2Mr^2+36(\Lambda-2)M^2r+72M^3}
	{r^3\left[(\Lambda-2)r+6M\right]^2}\ .
\end{equation}
The even-parity matter-source has a rather extended expression given by
\begin{eqnarray}
\label{source:polar}
\fl
S^{({\rm e})}=-\frac{8\pi}{\Lambda\left[(\Lambda-2)r+6M\right]}
\Bigg\{
\frac{\Lambda\Big(6r^3-16Mr^2\Big)-r^3\Lambda^2-8r^3+68Mr^2-108M^2r}
     {(\Lambda-2)r+6M} T_{00}^{{\ell m}}
\nonumber\\
\fl\hskip 2.25cm
	+\frac{1}{e^{4b}}\Big[2Mr+r^2(\Lambda-4)\Big]T_{11}^{{\ell m}}+
	2r^3 \partial_{r_*} T_{00}^{{\ell m}}
	-\frac{2 r^3}{e^{4b}} 
	\partial_{r_*} T_{11}^{{\ell m}}
	\nonumber\\ 
\fl\hskip 2.25cm
	+\frac{4\Lambda r}{e^{4b}}T_{1}^{{\ell m}}
	+\frac{1}{e^{2b}}\left[2\Lambda\left(1-\frac{3M}{r}\right)-
	  \Lambda^2\right]T_{2}^{{\ell m}}+
	  \frac{4r^2}{e^{4b}}T_3^{{\ell m}}\Bigg\}
	\ .
\end{eqnarray}
and we refer the interested reader to~\cite{martel04, nagar04, nagar05a} 
where the steps necessary for the derivation of equation~(\ref{source:polar}) 
are given in detail and some application to accretion problems are discussed. 
Here, we simply recall that the expressions of the even-parity vector and tensor 
spherical-harmonics for the matter-source needed in (\ref{source:polar}) can be 
obtained from the orthogonality properties of the harmonics and are
\begin{eqnarray}
\fl && T_A^{{\ell m}}=\frac{1}{\Lambda}\int d\Omega 
	\left[t_{A 2} \partial_{\theta}(Y^*_{\ell m}) -
	t_{A 3} \frac{{\rm i}m Y^*_{\ell m}}{\sin^2\theta} \right]\ ,
	\hskip 2.7cm A = 0,1\ ,\\
\fl && T_{2}^{{\ell m}}=\frac{2}{\Lambda(\Lambda-2)} \int
	d\Omega\bigg[t_{22} \frac{W^*_{\ell m}}{2}  + 
	t_{23} \frac{2X^{*}_{{\ell m}}}{\sin\theta} 
	+t_{33} \bigg(\frac{\Lambda Y^*_{{\ell m}}}{2} 
	-\frac{m^2 Y^*_{{\ell m}}}{\sin^2\theta} +
	\cot\theta \partial_{\theta}Y^*_{\ell m}\bigg)
	\bigg]\ , 
	\nonumber\\\fl && ~\\
\fl && T_{3}^{{\ell m}} =\frac{1}{2r^2}\int d\Omega 
	\left(t_{22}+t_{33}\frac{1}{\sin^2\theta}\right)Y^*_{\ell m}\ , \\
\fl && T_{AB}^{{\ell m}}=\int d\Omega\,t_{AB}\, Y^*_{\ell m}\ ,\;
	\hskip 6.0cm A,B = 0,1 \ ,
\end{eqnarray}
where the angular functions $W^{\ell m}(\theta, \phi)$ and $X^{\ell
m}(\theta, \phi)$ are defined as~\cite{rw57}
\begin{eqnarray}
\label{Wlm}
\fl W^{{\ell m}}
	&\equiv\frac{\nabla_{(\phi}S_{\theta)}^{\ell m}}{\sin\theta}=
\partial^2_{\theta}Y^{{\ell m}}-\cot\theta\,\partial_{\theta}
	Y^{{\ell m}}-\frac{1}{\sin^2\theta}\partial^2_{\phi}
	Y^{{\ell m}}\ ,\\ 
\label{Xlm}
\fl X^{{\ell m}}
	&\equiv-\sin\theta\left(\nabla_{\theta}S_{\theta}^{\ell m}-
	\frac{\nabla_\phi
	S_{\phi}^{\ell m}}{\sin^2\theta}\right)= 2\left(\partial^2_{\theta\phi}
	Y^{{\ell m}}-\cot\theta\partial_{\phi}Y^{{\ell m}}\right)\ .
\end{eqnarray}
and where the asterisk stands for complex conjugation. A few comments 
are now needed about the different notations in which the even-parity 
functions can be found in the literature. A particularly common choice 
is that proposed by Moncrief in~\cite{moncrief74} for the the even-parity 
gauge-invariant master function $Q^{({\rm e})}$ which is related to 
Zerilli function (\ref{psi_e}) simply as 
$Q^{({\rm e})} = \Lambda \Psi^{({\rm e})}$. 
Other authors (e.g.~\cite{martel04,lousto97}) use instead a master 
function defined as $Z\equiv 2{\Psi}^{({\rm e})}$ and a factor of 2 
needs to be introduced or removed when converting between the two 
notations. Yet another even-parity function can be introduced in terms 
of two new gauge-invariant metric functions $k$ and $\chi$ 
where~\cite{gundlach00,gundlach01}
\begin{eqnarray}
\fl && k = \kappa_1 =
	K+\frac{1}{e^{2b}}\left[r \partial_r G-
	\frac{2}{r}h_1^{({\rm e})}\right]\ ,
\\
\fl && \chi + k = H_2 -\frac{2}{e^{2b}}\de_rh_1^{(\rm e)}-
	\frac{2M}{r^2}h_1^{(\rm e)} 
        + \frac{1}{e^{2b}}\de_r\left(r^2\de_rG\right)+M\de_r G\ ,
\end{eqnarray}
and such that
\begin{equation}
\kappa_2 = \frac{1}{2}e^{2b}\left(\chi-r\de_r k+\frac{M}{r}e^{2b} k\right)\ .
\end{equation}
In this case, the Zerilli function (\ref{psi_e}) can be equivalently
defined as~\cite{gundlach01}
\begin{equation}
\label{psi_e_gmg}
{\Psi}^{({\rm e})}\equiv \frac{2r^2}
	{\Lambda\left[(\Lambda-2)r+6M\right]e^{2b}}
	\left[\chi+\left(\frac{\Lambda}{2}+\frac{M}{r}\right)e^{2b} 
	k -r \partial_{r} k\right]\ .
\end{equation}

	Finally, we recall that the homogeneous odd and even-parity
master equations (\ref{rw_eq}) and (\ref{zerilli_eq}) can be transformed
into each other by means of differential operations~\cite{chandra} and
they are connected to the master equation that Bardeen and
Press~\cite{BP_1973} have derived via the Newman-Penrose formalism.

\subsection{Initial data}
\label{id}

	An important aspect in the solution in the time-domain of the
perturbation equations (\ref{rw_eq}) and (\ref{zerilli_eq}) is the
specification of the initial data, namely the determination of the
Regge-Wheeler and Zerilli functions $\Psi^{(\rm o)}$ and $\Psi^{(\rm e)}$
at a given initial time and in the presence of the external sources of
perturbations. The standard approach in this case is that of exploiting
the $1+1$ splitting of spacetime and to solve the perturbed Hamiltonian
and momentum-constraint equations as a set of coupled ordinary
differential equations having the matter perturbations as sources.

	In the case of even-parity perturbations, in particular, the
Hamiltonian constraint takes the form~\cite{martingarcia99,gundlach00}
\begin{equation}
\label{const:ham}
\fl  \de^2_{r_*}k + \frac{2- 5M}{r} \de_{r_*}k
	-\frac{r-2M}{r^2}\de_{r_*}\chi
	-\frac{r-2M}{r^3}\left[\Lambda k-\left(\frac{\Lambda}{2}
	+1\right)\chi\right]=-8\pi T^{\ell m}_{00}\ ,
\end{equation}
and the momentum constraint is instead given by
\begin{eqnarray}
\label{const:mom_even}
\fl \de^2_{r_*}{\dot k}+\frac{2(r-2M)}{r^2}\de_{r_*}{\dot k} - 
	\frac{2M^2+r\Lambda(r-2M)}{2r^4}{\dot k}
	+\left(1-\frac{2M}{r}\right)\frac{4(r-2M)-
	\Lambda r}{4r^3}{\dot \chi} 
\nonumber\\
\fl \hskip 1.25cm -\frac{r-2M}{r^2}\de_{r_*}{\dot \chi}
	=\frac{4\pi}{r-2M}\left\{4r\de_{r_*}T^{\ell m}_{01}+\frac{1}{r}
	\left[\Lambda T_0^{\ell m}+4\left(2r-3M\right)
	T_{01}^{\ell m}\right]\right\}\ ,
\end{eqnarray}
where ${\dot k}\equiv \partial_t k$ and ${\dot \chi}\equiv \partial_t
\chi$. 

	Clearly, equations (\ref{const:ham}) and (\ref{const:mom_even})
are not sufficient to determine all of the unknowns: i.e. ($k,
{\dot k}$) and ($\chi, {\dot \chi}$) and this degree of indeterminacy
reflects the arbitrary amount of gravitational radiation that can be
present at the initial time with the same matter-sources. To alleviate at
least in part this problem, the initial data is chosen so as to be best
adapted to the physical problem under consideration. To minimize the
initial gravitational-wave content, for instance, it is possible to set
$\chi = 0 ={\dot \chi}$ (conformally-flat approximation) and to use
eqs. (\ref{const:ham}) and (\ref{const:mom_even}) to compute $k$ and
${\dot k}$. Alternatively, it is possible to consider time-symmetric
initial data for which ${\dot k} = 0 = {\dot \chi}$ and compute $\chi$
and $k$ from the constraint equations~\cite{martel02}. Finally, it is
possible to specify a solution for $\chi$ and ${\dot \chi}$ and use the
constrain equations to compute $k$ and ${\dot k}$~\cite{ruoff01b}.

	Similarly, in the case of odd-parity perturbations, we can
exploit the linearized Einstein equations [cf equations~(18) 
and~(20) of~\cite{martingarcia99}]
\begin{eqnarray}
\label{odd_constr_eq1}
&&g{\!\!\!\!\; \raisebox{-0.1ex}{$^{^0}$}}^{CD}\nabla_{\!D} 
	\left[4r(\nabla_{\![A} r) k_{C]} + 
	2 r^2 \nabla_{\![C} k_{A]}\right]
	- (\Lambda-2) k_{A} = -16\pi L^{\ell m}_{A} \ ,
\\\nonumber \\ 
\label{odd_constr_eq2}
&&\nabla_{\!A}k^A=16\pi L^{\ell m}\ ,
\end{eqnarray}
where
\begin{equation}
\label{multipole_L}
\fl L^{\ell m} \equiv \frac{1}{\Lambda(\Lambda-1)}\int\frac{d\Omega}
	{\sin\theta}\left[t_{23}W^*_{\ell m}
	+\frac{1}{2}\left(-t_{22}+\frac{1}{\sin^2\theta}t_{33}\right)
	X^*_{\ell m}\right]\;.
\end{equation}
Tedious but straightforward algebra allows to select two of the three
equations~(\ref{odd_constr_eq1}) and (\ref{odd_constr_eq2}) as constraint
equations for the four unknowns ($k_0, {\dot k}_0$) and ($k_1, {\dot
k}_1$)
\begin{eqnarray}
\label{id_odd1}
\fl \de_{r_*}^2k_1+\frac{2M}{r^2}\de_{r_*}k_1-\frac{4M}{r^2 e^{4b}}
	k_1-e^{2b}\de_{r_*}\dot{k}_0 +\frac{2M}{r^2}e^{2b}\dot{k}_0 = 16\pi\de_{r_*}L^{\ell m} \\
\nonumber\\
\label{id_odd2}
\fl \de^2_{r_*}k_0+\frac{2(r-3M)}{r^2}\de_{r_*}k_0+
	\left(\frac{2M}{r^3}-\frac{\Lambda}{r^2}\right)\frac{1}{e^{2b}}k_0+
	\frac{2M}{r^3 e^{4b}}k_1 -\frac{2}{r e^{2b}}\dot{k}_0
\nonumber\\
\hskip 4.0cm
	-\frac{2}{r e^{4b}}\dot{k}_1
	-\frac{1}{e^{2b}}\de_{r_*}\dot{k}_1
	=-\frac{16\pi}{e^{2b}}L_0^{\ell m} \ ,
\end{eqnarray}
where ${\dot k_A}\equiv \partial_t k_A$. As for the even-parity
pertubations, the constraint equations (\ref{id_odd1}) and
(\ref{id_odd2}) can be solved as a set of coupled ordinary differential
equations to compute two of the four unknowns when the other two are
specified according to the problem under consideration.

	As a final remark we recall that an equivalent and alternative
approach to the specification of the initial data consist in setting
$\partial_t \Psi^{({\rm o})}=0=\partial_t \Psi^{({\rm e})}$ and in
solving the Regge-Wheeler and Zerilli equations as ordinary differential
equations~\cite{CPM78,CPM79}. In the case of even-parity perturbations
this is equivalent to setting ${\dot k} = 0 = {\dot \chi}$ 
(cf equation~(\ref{psi_e_gmg})) and thus considering 
time-symmetric initial data.

\subsection{Frequency domain and QNMs}
\label{sec:fd}

	We conclude this Section by recalling that the investigation of
the Regge-Wheeler and Zerilli equations (\ref{rw_eq})--(\ref{zerilli_eq})
is often made in the frequency domain, where the the properties of
Fourier transform can be exploited and the oscillation properties of the
spacetime are easier to interpret. What is usually done in this case is
to assume a harmonic time dependence in the perturbations, i.e.
$\Psi^{({\rm e})} \sim \exp (i \omega_n t)$ and $\Psi^{({\rm o})} \sim \exp
(i \omega_n t)$ where $\omega_n$ is the complex oscillation frequency of
the $n$-th mode, so that the Regge-Wheeler and Zerilli equations
(\ref{rw_eq})--(\ref{zerilli_eq}) take respectively the form
\begin{eqnarray}
\label{rw_fd}
&&\left(\Psi^{({\rm o})}\right)^{''} + \left( \omega^{2}_n-
	V^{({\rm o})}_{\ell} \right) \Psi^{({\rm o})} = S^{({\rm o})} \ ,
\\ 
\label{zm_fd}
&&\left(\Psi^{({\rm e})}\right)^{''} + \left( \omega^{2}_n-        
	V^{({\rm e})}_{\ell} \right) \Psi^{({\rm e})} = S^{({\rm e})}  \ ,
\end{eqnarray}
where a `prime' indicates a total derivative with respect to
$r_*$. Equations (\ref{rw_fd})--(\ref{zm_fd}) can be solved as an
eigenvalue problem with boundary conditions such that the perturbations
behave as {\sl pure outgoing-waves} at spatial infinity 
\begin{equation}
\Psi^{({\rm o})}, \ \ \Psi^{({\rm e})} \sim \exp(-i \omega_n r_*) 
	\hskip 1.0cm {\rm for}\ \ r_*\rightarrow + \infty \ .
\end{equation}
and as {\sl pure ingoing-waves} at the event horizon
\begin{equation}
\Psi^{({\rm o})}, \ \ \Psi^{({\rm e})} \sim \exp(i  \omega_n r_*) 
	\hskip 1.3cm {\rm for}\ \ r_*\rightarrow -\infty\ ,
\end{equation}
In the absence of source terms, the modes solutions of 
equations~(\ref{rw_fd}) and (\ref{zm_fd}) are referred to as the 
quasi-normal modes (QNMs) of the black hole, with the real part of 
$\omega_n$ representing the oscillation frequency and with the 
imaginary part of $\omega^{-1}_n$ representing the damping time of 
the oscillation. A complete discussion of the properties and 
astrophysical relevance of these modes can be found 
in~\cite{chandra, frolov98, ks99} and here we only recall the
most important properties for a Schwarzschild black hole:

\begin{itemize}

\item All of the QNMs have positive imaginary parts and represent
  therefore damped modes. Stated differently, a Schwarzschild black hole
  is {\sl linearly stable} against perturbations.

\item The damping time of QNMs depends linearly on the mass of the black
  hole and is shorter for higher-order modes.

\item The excitation of a black hole is referred to as its ``ringing''
  and the late-time dynamics of these perturbations (i.e. the
  `tail' of the ringing) can be described with a power-law representing
  the envelope of the various decaying QNMs.

\item The QNMs are {\it isospectral}, i.e. odd and even-parity
  perturbations have the same complex eigenfrequencies. This is due to
  the uniqueness in which a black hole can react to a perturbation and it
  is not true for a relativistic star.

\end{itemize}

\section{Asymptotic expressions and gravitational waves}
\label{ae}

	In the previous Section we have reviewed the derivation of the
equations describing the evolution of perturbations of nonrotating black
holes induced by a non-zero stress-energy tensor. These perturbations have
been assumed to be generic in nature, needing to satisfy only the
condition of having a mass-energy much smaller than that of the black
hole. The solution of these equations with suitable initial conditions
completely specifies the reaction of the black hole to the perturbations
and this is essentially represented by the emission of gravitational
waves.

	As mentioned in Section \ref{gi_intro}, the importance of the
gauge-invariant variables used so far is that they are directly related
to the amplitude and energy of the gravitational-wave signal measured at
large distances. The purpose of this Section is to review the steps
necessary to obtain the relations between the master functions for the
odd and even-parity perturbations and the `plus' and `cross'
polarization amplitudes $h_+, h_\times$ of a gravitational wave in a
transverse and traceless (TT) gauge. In practice, and following the
guidelines tracked in~\cite{CPM78,CPM79}, we will derive an
expression for the perturbation metric ${\boldsymbol h}$ equivalent to
that obtained in the standard TT-gauge on a Minkowski spacetime and
relate it to the odd and even-parity master functions $\Psi^{({\rm o})}$
and $\Psi^{({\rm e})}$.

	To obtain this result a number of conditions need to be
met. Firstly, we need to evaluate each multipole of the decomposed metric
perturbations in the tetrad ${\mathbf e}$ of stationary observers in the
background Schwarzschild spacetime, i.e. $h_{\hat{\mu}\hat{\nu}} =
{\mathbf e}^{\mu}_{\;\hat{\mu}} {\mathbf e}^{\nu}_{\;\hat{\nu}}
h_{\mu\nu}$, where ${\mathbf e}$ is diagonal with components ${\mathbf
e}_{\;\hat{\mu}}^{\mu}\equiv \left \{e^b, e^{-b}, r^{-1}, (r
\sin\theta)^{-1}\right\}$ and where the indices ${\hat \mu}$ refer to the
locally `flat' coordinates. Secondly, all of the quantities need to be
evaluated far away from the source (i.e. in the ``wave-zone'') and
in the so-called {\it radiation gauge}. In practice, this amounts to
requiring that components $h_{\hat{\theta}\hat{\theta}}$,
$h_{\hat{\phi}\hat{\phi}}$ and $h_{\hat{\theta}\hat{\phi}}$ are functions
of the type $f(t-r)/r$ (i.e. they are outgoing spherical waves),
while all the other components have a more rapid decay of ${\cal
O}(1/r^2)$.  Thirdly, we need to impose the condition that the metric is
traceless modulo higher order terms, i.e. $h_{\hat\theta\hat\theta}
+ h_{\hat\phi\hat\phi}= 0 + {\cal O}(1/r^2)$.

\subsection{Asymptotic expressions from odd-parity perturbations}
\label{sec:h_from_op}

	We first consider odd-parity perturbations and recall that from
the radiation-gauge conditions and since for large $r$ the metric
asymptotes that of a flat spacetime $e^{b}\sim e^{-b}\sim 1$, we have
\begin{eqnarray}
\label{label:h0}
h_{\hat{\theta}\hat{t}}^{({\rm o})}& =
	\frac{h_0^{({\rm o})}}{r}e^{b}S_{\theta}\sim 
	\frac{h_0^{({\rm o})}}{r}\sim
	{\cal O}\left(\frac{1}{r^2}\right)
	\;\longrightarrow h_0^{({\rm o})}\sim
	{\cal O}\left(\frac{1}{r}\right)\ ,\\
\label{label:h1}
h_{\hat{\theta}\hat{r}}^{({\rm o})}& =
	\frac{h_1^{({\rm o})}}{r}e^{b}S_{\theta}\sim 
	\frac{h_1^{({\rm o})}}{r}\sim {\cal O}\left(\frac{1}{r^2}\right)
	\;\longrightarrow h_1^{({\rm o})}\sim 
	{\cal O}\left(\frac{1}{r}\right)\ ,
\end{eqnarray}
where the $\ell, m$ indices have been omitted for clarity.  Similarly,
since $h_{\hat{\theta}\hat{\theta}}^{({\rm o})} = 2h r^{-2}
\nabla_{\theta} S_{\theta}\sim {\cal O}(1/r)$ we deduce that $h\sim {\cal
O}(r)$ so that the only components of the metric having wave-like
properties at large $r$ are
\begin{eqnarray}
\label{odd:hplus}
h_{+}^{({\rm o})} &\equiv
	&\frac{1}{2}\left(h_{\hat{\theta}\hat{\theta}}^{({\rm
	o})}-h_{\hat{\phi}\hat{\phi}}^{({\rm o})}\right)
	=\frac{h}{r^2}\left(\nabla_{\theta}S_{\theta}-
	\frac{\nabla_{\phi}S_{\phi}}{\sin^2\theta}\right)+
	{\cal O}\left(\frac{1}{r^2}\right)\ , \\
\label{odd:hcross}
h_{\times}^{({\rm o})}&\equiv & h_{\hat{\theta}
	\hat{\phi}}^{({\rm o})}=\frac{h}{r^2}
	\frac{\nabla_{(\phi}S_{\theta)}}{\sin\theta}+
	{\cal O}\left(\frac{1}{r^2}\right)\ ,
\end{eqnarray}
and where we note that since $h$ has the dimensions of a length squared,
both $h_{+}$ and $h_{\times}$ are, as expected, dimensionless.

	Next, we need to relate the perturbation $h$ to the odd-parity
master function $\Psi^{({\rm o})}$. To do so, we follow the procedure
outlined in~\cite{CPM78}, and note that (cf equation~(III-20)
\footnote{We recall that in the notation of~\cite{CPM78}
$\widetilde{\psi} = \Lambda(\Lambda-2)\Psi^{(\rm o)}$, and the multipoles
in~\cite{CPM78} are related to ours as $\widetilde{h}_2 = h_2 = -2h$,
$\widetilde{h}_{0} = h^{(\rm o)}_{0}$ and $\widetilde{h}_{1} = h^{(\rm
o)}_{1}$})
\begin{equation}
\label{eq:III-20}
  \de_t h =\left(1-\frac{2M}{r}\right)\de_r\left(r\Psi^{(\rm o)}\right)
  +h_0^{(\rm o)}\ ,
\end{equation}
Equation~(\ref{eq:III-20}) represents one of the Hamilton equations as
derived by Moncrief in a Hamiltonian formulation of perturbation
equations~\cite{moncrief74} (see~\ref{appendix_d} for details). We also
note that the radiation-gauge conditions on $h$ and $h_0^{(\rm o)}$ imply
that $\Psi^{({\rm o})}\sim {\cal O}(1)$, i.e., in the wave-zone
$\Psi^{({\rm o})}$ has the dimensions of a length, behaves as an
outgoing-wave, but it does not depend explicitly on $r$.

	Exploiting now the outgoing-wave behaviour of $h$ at large
distances we can write
\begin{equation}
\label{h_outgoing}
\de_t h=-\de_r h+{\cal O}\left(\frac{1}{r}\right)\ ,
\end{equation}
so that asymptotically equation~(\ref{eq:III-20}) simply becomes
\begin{equation}
\de_r h=-\de_r\left(r\Psi^{(\rm o)}\right)+{\cal O}\left(\frac{1}{r}\right)\ ,
\end{equation}
and its integration yields
\begin{equation}
\frac{h}{r}\sim -\Psi^{({\rm o})}+{\cal O}\left(\frac{1}{r}\right)\ .
\end{equation}
As a result, the `$+$' and `$\times$' polarization amplitudes of the
gravitational wave can be calculated from
eqs. (\ref{odd:hcross})--(\ref{odd:hplus}) as
\begin{eqnarray}
\label{odd:plus_b}
h_{+}^{({\rm o})} &
	=-\frac{1}{r}\Psi^{({\rm o})}\;
	\left(\nabla_\theta S_{\theta}-
	\frac{\nabla_\phi S_{\phi}}{\sin^2\theta}\right)+
	{\cal O}\left(\frac{1}{r^2}\right)\ , \\
\label{odd:cross_b}
h_{\times}^{({\rm o})}&
	=-\frac{1}{r}
	\Psi^{({\rm o})}\;
	\frac{\nabla_{(\phi}S_{\theta)}}
	{\sin\theta}+{\cal O}\left(\frac{1}{r^2}\right)\ .
\end{eqnarray}

	It is now interesting to note that expressions (\ref{odd:plus_b})
and (\ref{odd:cross_b}) can be written in a compact form using the $s=-2$
spin-weighted spherical harmonics
\begin{equation}
\label{eq:spin_harmonic}
_{_{-2}}Y^{{\ell m}}(\theta,\phi)\equiv\sqrt{\frac{(\ell-2)!}{(\ell+2)!}}
	\left(W^{{\ell m}}-\mathrm{i}\;
	\frac{X^{{\ell m}}}{\sin\theta}\right)\ ,
\end{equation}
so that expressions (\ref{odd:plus_b}) and (\ref{odd:cross_b}) can be
combined into a single complex expression given by
\begin{equation}
\label{h+_hx_Psio}
\left(h^{({\rm o})}_+ -
	\mathrm{i}h^{({\rm o})}_{\times}\right)_{{\ell m}}=
	\frac{\mathrm{i}}{r}\;\sqrt{\frac{(\ell+2)!}{(\ell-2)!}}\;
	\Psi^{({\rm o})}_{{\ell m}}\;_{_{-2}}Y^{{\ell m}}(\theta,\phi) 
	+ {\cal O}\left(\frac{1}{r^2}\right) \ ,
\end{equation}
where, for clarity, we have explicitly restored the multipole indices
$\ell, m$.

	An important clarification is now needed. We have discussed in
the previous Section that the odd-parity metric perturbations are
sometimes expressed in terms of the odd-parity Moncrief function
$Q^{({\rm o})}$ (cf equation~(\ref{mfop})] and it is not unusual to
find in the literature the gravitational wave amplitudes expressed in
terms of this quantity. However, great care must be paid to the
asymptotic relation between the master function $Q^{({\rm o})}$ and the
gravitational-wave amplitudes and, indeed, this is sometimes source of
confusion~\cite{ko04,kon03}. To clarify this point, we recall that the
derivation of the asymptotic relation between $Q^{({\rm o})}$ and $h$
proceeds as in a way similar to the one discussed above. In the radiation
gauge and at large distances from the black hole, we can use relation
(\ref{h_outgoing}) in the definition (\ref{mfop}) with $h_2=-2h$, so that
\begin{equation}
\label{Q(o)_vs_h}
Q^{({\rm o})}\sim \frac{1}{r} \partial_t h + 
	{\cal O}\left(\frac{1}{r}\right)\ ,
\end{equation}
which is also a dimensionless quantity. Since $h \sim {\cal O}(r)$, the
function $Q^{({\rm o})}$ does not depend on $r$ at leading order and
equation~(\ref{Q(o)_vs_h}) can be integrated to give
\begin{equation}
\label{ht_over_r}
\frac{h(t)}{r}\sim \int_{-\infty}^{t} Q^{({\rm o})}(t') dt' 
	+ {\cal O}\left(\frac{1}{r}\right) + \mathrm{const.}\ ,
\end{equation}
where the integration constant can be defined as 
\begin{equation}
\mathrm{const.}\equiv \lim_{t\rightarrow - \infty} \frac{h(t,r)}{r}\sim
	{\cal O}(1)\ ,
\end{equation}
and it can be set to zero in the case of asymptotically flat
metric perturbations ($h=0$) at earlier times. Combining now expressions
(\ref{Wlm}), (\ref{Xlm}) and (\ref{ht_over_r}), the gravitational-wave
amplitudes in the two polarizations and with the new master function read
\begin{equation}
\label{h+_hx_Qo}
\fl \left(h^{({\rm o})}_+
	-\mathrm{i}h^{({\rm o})}_{\times}\right)_{{\ell m}}=
	-\frac{\mathrm{i}}{r}\sqrt{\frac{(\ell+2)!}{(\ell-2)!}}
	\left(\int_{-\infty}^{t} Q^{({\rm o})}_{{\ell m}}(t')dt'
	\right)\;_{_{-2}}Y^{{\ell m}}(\theta,\phi)+
	{\cal O}\left(\frac{1}{r^2}\right)\ .
\end{equation}

	Comparing expressions (\ref{h+_hx_Psio}) and (\ref{h+_hx_Qo}) we
note that while $\Psi^{({\rm o})}$ and $Q^{({\rm o})}$ are both solutions
of the Regge-Wheeler equation (\ref{rw_eq}), they yield two different
asymptotic expressions for the gravitational-wave amplitudes. This
difference, which is consistent with equation~(\ref{Q_vs_Psi}) when evaluated
in a an asymptotic region of the spacetime where $L_1^{\ell m}=0$, is
subtle but important and, as mentioned above, it has led to some
inconsistencies in the literature both for the determination of the
asymptotic gravitational-wave amplitudes and for the energy losses. This
will be further discussed in sections~\ref{a_general} and
\ref{e_am_losses}.

\subsection{Asymptotic expressions from even-parity perturbations}
\label{sec:h_from_ep}

	A calculation conceptually analogous to the one carried out in
sections~\ref{sec:h_from_op} leads to the relation between the
gravitational-wave amplitude and the even-parity master function. In
particular, after projecting equation~(\ref{even:metric}) along the stationary
tetrad, the asymptotic wave amplitudes in the two polarization states are
\begin{eqnarray}
\fl h_{+}^{({\rm e})}&=\frac{1}{2}
	\left(h_{\hat{\theta}\hat{\theta}}^{({\rm e})}-
	h_{\hat{\phi}\hat{\phi}}^{({\rm e})}\right)=
	\frac{G}{2}\left(\nabla_\theta\nabla_\theta
	Y^{{\ell m}}-\frac{\nabla_\phi\nabla_\phi
	Y^{{\ell m}}}{\sin^2\theta}\right)=\frac{G}{2}\;W^{{\ell m}}\ ,\\ 
\fl h_{\times}^{({\rm e})}&=h_{\hat{\theta}\hat{\phi}}^{({\rm
	e})}=G\,
	\frac{\nabla_\theta\nabla_\phi Y^{{\ell m}}}{\sin\theta}
	=\frac{G}{2}\;\frac{\;\;\;X^{{\ell m}}}{\sin\theta}\ ,
\end{eqnarray}
so that we essentially need to relate the metric perturbation $G$ with the
even-parity function ${\Psi}^{({\rm e})}$. Firstly, it is easy to realize
that the even-parity metric projected onto the tetrad,
$h_{\hat\mu\hat\nu}^{({\rm e})}$ is such that 
\begin{equation}
H_2\sim {\cal O}\left(\frac{1}{r^2}\right)\ ,  \qquad {\rm and} \qquad 
h_{1}^{({\rm e})} \sim {\cal O}\left(\frac{1}{r}\right) \ , 
\end{equation}
so that the terms proportional to these multipoles are of higher order
for large $r$ and can be neglected. Furthermore, from the traceless
condition 
\begin{equation}
h_{\hat{\theta}\hat{\theta}}^{({\rm e})}+
	h_{\hat{\phi}\hat{\phi}}^{({\rm e})}= 0 + 
	{\cal O}\left(\frac{1}{r^2}\right) \ ,
\end{equation}
we obtain an asymptotic relation between $K$ and $G$ of the type
\begin{equation}
\label{K_vs_G_1}
\fl 2KY^{{\ell m}}+G\left(\nabla_\theta\nabla_\theta
	Y^{{\ell m}}+\frac{\nabla_\phi\nabla_\phi 
	Y^{{\ell m}}}{\sin^2\theta}\right)
	=\left(2K-G\Lambda\right)Y^{{\ell m}}\sim 
	{\cal O}\left(\frac{1}{r^2}\right)\ ,
\end{equation}
where we have used the definition (\ref{def:harmonics}) to derive the
right-hand-side of expression (\ref{K_vs_G_1}). As a result, the
asymptotic relation between the two components of the even-parity part of
the perturbation metric simply reads
\begin{equation}
K\sim\frac{\Lambda}{2}\,G+{\cal O}\left(\frac{1}{r^2}\right)\ .
\end{equation}
Using now the definitions (\ref{kappa1})--(\ref{kappa2}), we have that
asymptotically
\begin{eqnarray}
\kappa_1&\sim& \frac{\Lambda}{2}\,G + r \partial_r G+
	{\cal O}\left(\frac{1}{r^2}\right)\\ \kappa_2&\sim&
	-\frac{1}{2}\left(K+r \partial_r K\right)\sim
	-\frac{\Lambda}{4}\left(G+r \partial_r G\right)+
	{\cal O}\left(\frac{1}{r^2}\right)\ ,
\end{eqnarray}
and their linear combination~(\ref{def:q1}) becomes
\begin{equation}
q_1\sim \frac{rG}{2}\;\Lambda\left(\Lambda-2\right)+ 
	{\cal O}\left(\frac{1}{r}\right)\ .
\end{equation}

	Finally, the asymptotic gauge-invariant even-parity master
function reads
\begin{equation}
{\Psi}^{({\rm e})}\sim \frac{r q_1}{\Lambda\left[r(\Lambda-2)+
	6M\right]}\sim \frac{1}{2}\,rG +{\cal O}\left(\frac{1}{r}\right)\ ,
\end{equation}
so that, modulo higher-order terms, the even-parity gravitational-wave
amplitudes measured by a distant observer can be written in the compact
form
\begin{equation}
\left(h^{({\rm e})}_{+}-\mathrm{i}h^{({\rm e})}_{\times}\right)_{{\ell m}}=
	\frac{1}{r}\sqrt{\frac{(\ell+2)!}{(\ell-2)!}} 
	{\Psi}^{({\rm e})}_{{\ell m}}\;_{_{-2}}Y^{{\ell m}}(\theta,\phi)+
	{\cal O}\left(\frac{1}{r^2}\right)\ .
\end{equation}

\subsection{Asymptotic general expressions}
\label{a_general}

	It is often convenient to combine the expressions for the
asymptotic gravitational wave amplitudes due to odd and even-parity
perturbations into a single expression of the type
\begin{equation}
\label{hp_hc}
\fl
h_{+}-\mathrm{i}h_{\times}=\frac{1}{r}\sum_{\ell,m}
	\sqrt{\frac{(\ell+2)!}{(\ell-2)!}}
	\left({\Psi}^{({\rm e})}_{{\ell m}}+
	\mathrm{i}{\Psi}_{{\ell m}}^{({\rm o})}\right)
	\;_{_{-2}}Y^{{\ell m}}(\theta,\phi)+
	{\cal O}\left(\frac{1}{r^2}\right)\ ,
\end{equation}
or, equivalently 
\begin{eqnarray}
\label{hp_hc_mf}
\fl h_+-\mathrm{i}h_{\times}=\frac{1}{r}\sum_{\ell,m}
	\sqrt{\frac{(\ell+2)!}{(\ell-2)!}}
	\left({\Psi}^{({\rm e})}_{{\ell m}}-\mathrm{i}
	\int_{-\infty}^tQ^{({\rm o})}_{{\ell m}}(t')dt'
	\right)\,_{_{-2}}Y^{{\ell m}}(\theta,\phi)
	+ {\cal O}\left(\frac{1}{r^2}\right)\ ,
\nonumber \\
\end{eqnarray}
where we have defined $h_{+}\equiv h^{({\rm o})}_{+} + h^{({\rm e})}_{+}$
and $h_{\times}\equiv h^{({\rm o})}_{\times} + h^{({\rm
e})}_{\times}$. Note that $X^{{\ell 0}}=0$ for any value of $\ell$, so
that in the case of axisymmetry the gravitational wave signal is
proportional to $W^{\ell 0}$ only.

	It is also useful to underline that while expression
(\ref{hp_hc}) resembles the corresponding expression (10) 
of~\cite{ko04}, it is indeed very different.  Firstly, 
because in~\cite{ko04} the Moncrief function is adopted for the odd-parity part
of the perturbations and hence, modulo a normalisation factor, the
function $\Psi^{({\rm o})}$ appearing there corresponds to our function
$Q^{({\rm o})}$ ( cf expression (\ref{mfop}) ). Secondly, because
with this choice for the odd-parity perturbations a time derivative is
needed in the asymptotic expression for the gravitational-wave amplitudes
(cf the discussion in the derivation of equation~(\ref{h+_hx_Qo})).
As a result, expression (10) of~\cite{ko04} (which is also missing
the distinction between the real and imaginary parts) should really be
replaced by our expression (\ref{hp_hc_mf}). A similar use of the
Moncrief function for the odd-parity part is present also 
in~\cite{stu_03,ss_03,ss_04}, where it is employed to calculate the
gravitational-wave content of numerically simulated spacetimes.

\subsection{Energy and angular momentum losses}
\label{e_am_losses}

	Using the expressions derived so far we can now estimate the
energy and angular momentum losses due to gravitational waves propagating
outwards to spatial infinity. More specifically, this can be done by
using expression (\ref{hp_hc}) and the definition of Isaacson's
stress-energy pseudo-tensor $\tau_{\mu \nu}$ for the gravitational-wave
field ${\boldsymbol h}$ propagating in the curved background field
${\boldsymbol g}{\!\!\!\!\;\raisebox{-0.1ex}{$^{^0}$}}$ and in a Lorentz
gauge~\cite{isaacson,landau}
\begin{equation}
\label{gw_set_c}
\tau_{\mu \nu} \equiv \frac{1}{32 \pi}
	\left\langle 
	\nabla_{\mu} {h}_{\alpha \beta} 
	\nabla_{ \nu} {h}^{\alpha \beta}
	\right\rangle \ ,
\end{equation}
where the brackets $\langle \ldots \rangle$ refer to a spatial average
over a length much larger than the typical gravitational
wavelength~\cite{isaacson,mtw74}. The averaged expression
(\ref{gw_set_c}) is gauge-invariant~\cite{isaacson} and holds in the
`limit of high frequency' (or {\it short-wave} approximation), i.e.  
whenever the wavelength of the gravitational-wave field is small
when compared to the local radius of curvature of the background
geometry. In practice, gravitational radiation from isolated systems is
of high frequency whenever it is far enough away from its source.

	Expression (\ref{gw_set_c}) accounts for the amount of energy and
momentum carried by the gravitational wave over a certain region of
spacetime, but since we are interested in the energy flux as measured by
an inertial observer, we need to project the pseudo-tensor on the
observer's locally orthonormal tetrad, where it becomes
\begin{equation}
\label{gw_set}
\tau_{\hat\mu \hat\nu} \equiv \frac{1}{32 \pi}
	\left\langle 
	\partial_{\hat\mu} {\bar h}_{\hat\alpha \hat\beta} 
	\partial_{\hat\nu} {\bar h}^{\hat\alpha \hat\beta}
	\right\rangle \ ,
\end{equation}
with $\bar{h}_{\hat\mu\hat\nu} \equiv h_{\hat\mu\hat\nu} - \frac{1}{2}h
\eta_{\hat\mu\hat\nu}$. As a result, the energy per unit time and angle
carried by the gravitational waves and measured by a stationary observer
at large distance from the black hole is given by
\begin{equation}
\label{dedtdom}
\fl\frac{d^2E}{dtd\Omega}=\frac{r^2}{16\pi}
	\left[ \left(\frac{d {h}_{\hat{\theta}\hat{\phi}}}{dt}\right)^2+
	\frac{1}{4}\left(\frac{d h_{\hat{\theta}\hat{\theta}}}{dt}-
	\frac{d h_{\hat{\phi}\hat{\phi}}}{dt}\right)^2\right]=
	\frac{r^2}{16\pi}\left(\left|\frac{d {h}_+}{dt}\right|^2
	+\left|\frac{d h_{\times}}{dt}\right|^2\right) \ , 
\end{equation}
where the total derivative is made with respect to the asymptotic
observer's time. Integrating (\ref{dedtdom}) over the solid angle, the
total power emitted in gravitational waves is then given by
\begin{eqnarray}
\label{power}
&& \frac{dE}{dt} =  
	\frac{1}{16\pi}\sum_{\ell,m}\frac{(\ell+2)!}{(\ell-2)!}
	\left(\left|\frac{d {\Psi}^{({\rm e})}_{{\ell m}}}{dt}\right|^2
	+\left|\frac{d {\Psi}^{({\rm o})}_{{\ell m}}}{dt}\right|^2\right) 
	\ , \\
\label{power:CPM}
&& \hskip 0.6cm	
	= \frac{1}{16\pi}\sum_{\ell,m} \left(
	\Lambda(\Lambda-2)
	\left|\frac{d {\Psi}^{({\rm e})}_{{\ell m}}}{dt}\right|^2
	+\frac{\Lambda}{\Lambda-2}
	\left|\frac{d {\Phi}^{({\rm o})}_{{\ell m}}}{dt}
	\right|^2\right)\ ,
\end{eqnarray}
where expression (\ref{power:CPM}) was first presented 
in~\cite{CPM78,CPM79}.

	For the same reasons discussed in the previous section, these
expressions need to be suitably modified when the energy losses are
expressed in terms of the odd-parity Moncrief function $Q^{({\rm o})}$,
in which case they assume the form
\begin{equation}
\frac{dE}{dt} = \frac{1}{16\pi}\sum_{\ell,m}\frac{(\ell+2)!}{(\ell-2)!}
	\left(\left|\frac{d {\Psi}^{({\rm e})}_{{\ell m}}}{dt}\right|^2+ 
	\left|Q_{{\ell m}}^{({\rm o})}\right|^2\right) \ .
\end{equation}

	 Besides having arbitrary angular dependence, the perturbations
can also have non-diagonal terms and thus account for a non-zero net
angular momentum, part of which can be lost to gravitational
radiation. In this case, the angular momentum flux carried away in the
form of gravitational waves can be calculated in terms of the $r \phi$
component of the stress energy tensor (\ref{gw_set}). In particular,
using spherical coordinates and assuming that the rotation is
parametrised by the angle $\phi$, we have
\begin{eqnarray}
\label{dJdtdom}
\frac{d^2J}{dtd\Omega}&=\frac{r^2}{32\pi}
	\left\langle \partial_{\phi}\bar{h}_{\hat{\mu}\hat{\nu}}
	\partial_r\bar{h}^{\hat{\mu}\hat{\nu}}\right\rangle \nonumber\\
	&= \frac{r^2}{16\pi}
	\left\langle\partial_rh_{\hat\theta\hat\phi}\partial_{\phi}
	h_{\hat\theta\hat\phi} +
	 \frac{1}{4}\partial_r\left(h_{\hat\theta\hat\theta}-h_{\hat\phi\hat\phi}\right)
 	\partial_\phi \left(h_{\hat\theta\hat\theta}-h_{\hat\phi\hat\phi}\right)
	\right\rangle \ .
\end{eqnarray}
Since $h_{\hat\theta\hat\theta}-h_{\hat\phi\hat\phi}=2h_+$
and $h_{\hat\theta\hat\phi}=h_{\times}$ and since the metric 
components with the radiation-gauge condition behave like 
outgoing spherical waves, so that $\de_r h_{+,\times}=-\de_t h_{+,\times}$, 
the angular momentum carried away in the form of gravitational 
waves (\ref{dJdtdom}) is then expressed as
\begin{equation}
\frac{d^2J}{dtd\Omega}=-\frac{r^2}{16\pi}\left(\partial_th_+\partial_\phi 
	h_+ + \partial_t h_{\times}\partial_{\phi}h_{\times}\right)\ .
\end{equation}
Proceeding in a way similar to the one followed in the calculation of the
emitted power, the total angular momentum lost per unit time to
gravitational wave reads~\cite{mp_05}
\begin{equation}
\label{dJdt_CPM}
\frac{dJ}{dt}=\frac{1}{32\pi}\sum_{\ell,m}\left\{{\rm i}m
	\frac{(\ell+2)!}{(\ell-2)!}
	\left[\frac{d\Psi^{(\rm e)}_{\ell m}}{dt}
	\left(\Psi_{\ell m}^{({\rm e})}\right)^*+
	\frac{d\Psi^{(\rm o)}_{\ell m}}{dt}
	\left(\Psi_{\ell m}^{({\rm o})}\right)^*\right]+c.c.\right\}\ ,
\end{equation}
or, using the Moncrief master function (\ref{mfop}) for the odd-parity
perturbations~\cite{poisson04} 
\begin{equation}
\label{dJdt_RWM}
\fl\frac{dJ}{dt}=\frac{1}{32\pi}\sum_{\ell, m}\left\{{\rm i} m
	\frac{(\ell+2)!}{(\ell -2)!}
	\left[\frac{d\Psi^{(\rm e)}_{\ell m}}{dt}
	\left(\Psi_{\ell m}^{({\rm e})}\right)^*+
	Q_{\ell m}^{(\rm o)}\int_{-\infty}^{t}
	\left(Q_{\ell m}^{(\rm o)}\right)^*(t')dt'\right]+c.c.\right\}\ .
\end{equation}

	To conclude, we report the expression for the energy spectrum
${dE}/{d\omega}$, which is readily calculated from equation~(\ref{power})
after performing the Fourier transform of the odd and even-parity master
functions, i.e.
\begin{equation}
\label{dedom}
\frac{dE}{d\omega}=\frac{1}{16\pi^2}\sum_{\ell,m}
	\frac{(\ell+2)!}{(\ell-2)!}\;\, \omega^2\left(
	\left|\widetilde{\Psi}^{({\rm e})}_{{\ell m}}\right|^2
	+\left|\widetilde{\Psi}^{({\rm o})}_{{\ell m}}\right|^2\right)\ ,
\end{equation}
where we have indicated with $\widetilde{f}(\omega,r)$ the Fourier
transform of the timeseries $f(t,r)$. As for the emitted power, also the
energy spectrum will have a different expression if the odd-parity
Moncrief function is used and in this case one obtains
\begin{equation}
\label{en_spectrum}
\frac{dE}{d\omega}=\frac{1}{16\pi^2}\sum_{\ell,m}
	\frac{(\ell+2)!}{(\ell-2)!}\;\,
	\left(\omega^2\left|\widetilde{\Psi}^{({\rm e})}_{{\ell m}}\right|^2
	+\left|\widetilde{Q}^{({\rm o})}_{{\ell m}}
	\right|^2\right) \ .
\end{equation}
	We note that expression (\ref{en_spectrum}) for the energy
spectrum agrees with the ones given in~\cite{ap99,lousto97} for the
cases of odd and even-parity perturbations, respectively and with the
complete expression in~\cite{martel04} for both odd and even-parity
multipoles. On the other hand, the energy spectrum reported 
in~\cite{ko04} suffers of the same inconsistencies discussed above for
equation~(\ref{hp_hc_mf}), so that equation~(27) in~\cite{ko04} should effectively
be replaced by equation~(\ref{en_spectrum}).

	We conclude this ection by noting that a new formalism has been
recently proposed to compute the energy and angular momentum fluxes
absorbed by a Schwarzschild (and Kerr) black hole in a form which is
suitable for time-domain computations~\cite{poisson04,mp_05}. Although
the effect of the absorbed flux is generally small in astrophysical
situations, it can be important in the secular dynamics in black-hole
spacetimes, as discussed in~\cite{martel04}. Referring the interested
reader to~\cite{poisson04,mp_05} for more details, here we simply
recall that the calculation of the absorbed energy flux cannot be made
with the present choice of Schwarzschild coordinates (in which the
ingoing flux at the horizon is suppressed by the infinite redshift) and
must therefore be performed using an advanced
time-coordinate. Furthermore, Isaacson's prescription for the
stress-energy pseudo-tensor (\ref{gw_set_c}) also ceases to be valid in
the vicinity of the horizon, where the short-wave approximation breaks
down and an alternative approach needs to be
implemented~\cite{poisson04,mp_05}.
	
\subsection{A commonly used convention}
\label{sec:RW}

	We conclude this section on asymptotic expressions by discussing
a rather popular choice for the gauge-invariant master functions and that
has found successful application in the extraction of the
gravitational-wave content of numerically simulated
spacetimes~\cite{abrahams96, bbhgca1, retal98, Rezzolla99a}. Furthermore,
the convention discussed below has been implemented in the {\tt Cactus}
computational toolkit~\cite{cs99, acs98}, a diffused and freely available
infrastructure for the numerical solution of the Einstein
equations~\cite{cactus}. Numerous tests and applications of this
implementation have been performed over the years and we refer the reader
to~\cite{cs99, acs98, fetal02, betal05} for examples both in vacuum
and non-vacuum spacetimes.

	The reference work for this convention in the one by Abrahams and
Price (1996)~\cite{abrahams96}, although a similar approach for the
even-parity part of the perturbations was also adopted in previous
works~\cite{abrahams92, anninos95b}. We first note that the coefficients
$c_0$, $c_1$ and $c_2$ introduced in~\cite{abrahams96} are related simply
to the multipolar coefficients of the odd-parity part introduced in
section~\ref{sec:multipole}. More specifically, $c_2 = - 2h = h_2$, $c_0
= h_0^{({\rm o})}$ and $c_1 = h_1^{({\rm o})}$ and it is then easy to
realise that the odd and even-parity master functions $Q^{\times}_{{\ell
m}}$ and $Q^{+}_{{\ell m}}$ defined in~\cite{abrahams96} are related to
the master functions discussed so far through the simple algebraic
expressions
\begin{eqnarray}
\label{qc}
&& Q^{\times}_{{\ell m}}=\sqrt{\frac{2(\ell+2)!}{(\ell-2)!}}\,
	Q^{({\rm o})}_{{\ell m}}\ , 
\\ 
\label{qp}
&& Q^{+}_{{\ell m}}=\sqrt{\frac{2(\ell+2)!}{(\ell-2)!}}\,
	{\Psi}^{({\rm e})}_{{\ell m}}\ ,
\end{eqnarray}
so that the asymptotic expression for the gravitational-wave amplitudes
in the two polarizations are given by
\begin{eqnarray}
\label{hp_ap}
\fl h_+=\frac{1}{\sqrt{2}r}\sum_{\ell,m}
	\sqrt{\frac{(\ell-2)!}{(\ell+2)!}}\,
	\left[Q^+_{{\ell m}}W^{{\ell m}}-\left(\int_{-\infty}^{t}
	Q^{\times}_{{\ell m}}(t')dt'\right)\frac{\;\;\;
	X^{{\ell m}}}{\sin\theta}\right]
	+ {\cal O}\left(\frac{1}{r^2}\right)\ , \\
\label{hc_ap}
\fl h_{\times}=\frac{1}{\sqrt{2}r}\sum_{\ell,m}
	\sqrt{\frac{(\ell-2)!}{(\ell+2)!}}\,
	\left[Q^+_{{\ell m}}\frac{\;\;\;X^{{\ell m}}}{\sin\theta}+
	\left(\int_{-\infty}^{t}Q^{\times}_{{\ell m}}(t')dt'\right)
	W^{{\ell m}}\right]
	+ {\cal O}\left(\frac{1}{r^2}\right)\ .
\end{eqnarray}
Also expressions (\ref{hp_ap}) and (\ref{hc_ap}) can be combined into a
single one
\begin{equation}
\fl h_{+}-\mathrm{i}h_{\times}=\frac{1}{\sqrt{2}r}\sum_{\ell,m}\left(
	Q^{+}_{{\ell m}} - \mathrm{i}\int_{-\infty}^{t}
	Q^{\times}_{{\ell m}}(t')dt'\right)
	\;_{_{-2}}Y^{{\ell m}}(\theta,\phi)
	+ {\cal O}\left(\frac{1}{r^2}\right)\ ,
\end{equation}
which closely resembles expression (\ref{hp_hc_mf}) and that in its
compactness highlights the advantage of the normalisation
(\ref{qc})--(\ref{qp}). Also very compact is the expression for the
emitted power that, with this convention, simply reads
\begin{equation}
\label{p_ap}
\frac{dE}{dt}=\frac{1}{32\pi}\sum_{\ell,m}\left(\left|
	\frac{d {Q}^{+}_{{\ell m}}}{dt}\right|^2+
	\left|Q^{\times}_{{\ell m}}\right|^2\right)\ .
\end{equation}
We note that expression (\ref{p_ap}) corrects equation~(24) 
of~\cite{abrahams96}, where a time derivative of the odd-parity
perturbations is present, leading to an obvious dimensional
inconsistency. This typo was subsequently corrected in~\cite{ap99}.

\section{Conclusions}
\label{conclusions}

	Black hole perturbation theory has been and will continue to be a
powerful tool with continuous developments and multiple applications. In
this paper we have briefly reviewed the theory of metric perturbations of
Schwarzschild black hole, concentrating in particular on its
gauge-invariant formulations. Special care has been paid to ``filter''
the technical details in the mathematical apparatus that may obscure the
important results and to ``steer'' the reader through the numerous
conventions and notations that have accompanied the development of the
formalism over the years, pointing out the misprints and errors that can
be found in the literature. Among the aims of this review is that of
providing the reader with a set of expressions in the two most common
formulations and that can be readily used for the calculation of the odd
and even-parity perturbations of a Schwarzschild spacetime in the
presence of generic matter-sources. In addition, we have reported the
asymptotic expressions for the gravitational-wave amplitudes and losses
driven by these generic perturbations. The latter can find useful
application in the calculation of the gravitational-wave content of
numerically simulated spacetimes.

	With suitable gauge choices, much of of what presented here for a
Schwarzschild black hole could be extended to a Kerr background and this
will be the subject of future work.

\ack
It is a pleasure to thank Andrew Abrahams, Tomohiro Harada, Frank
Herrmann, Vincent Moncrief, Eric Poisson, Richard Price, Ed Seidel
and Manuel Tiglio for useful discussions and comments. The activity of 
AN at the University of Valencia was supported by a ``Angelo della Riccia'' 
fellowship.

\appendix

\section{Some explicit expressions}

	In the general expressions given in the main text the tensor
harmonic appear in terms of covariant derivatives. While compact, such
expressions need to be `unwrapped' in practice and in what follows we
provide their explicit expressions.
\begin{eqnarray}
\nabla_{\theta}\nabla_{\theta}Y^{{\ell m}} &=& 
	\partial^2_{\theta}Y^{{\ell m}}\ ,\\
\nabla_{\phi}\nabla_{\phi}Y^{{\ell m}} &=&
	\partial^2_{\phi}Y^{{\ell m}}+
	\sin\theta\cos\theta\partial_{\theta}Y^{{\ell m}}\ ,\\
\nabla_{\phi}\nabla_{\theta}Y^{{\ell m}} &=&
	\partial^2_{\theta\phi}Y^{{\ell m}}-\cot\theta
	\partial_{\phi}Y^{{\ell m}}\ ,
\end{eqnarray}
for the even-parity ones, and
\begin{eqnarray}
S_{\theta}^{{\ell m}} &=& -\frac{1}{\sin\theta}\,
	\partial_{\phi}Y^{{\ell m}}\ ,
	\qquad \  S_{\phi}^{{\ell m}} = 
	\sin\theta\partial_{\theta}Y^{{\ell m}}\ ,\\
\nabla_{\theta}S_{\theta}^{{\ell m}} &=&
	-\frac{1}{\sin\theta}\left(\partial^2_{\theta\phi}
	Y^{{\ell m}}-\cot\theta\partial_{\phi}Y^{{\ell m}}\right)\ ,\\
\nabla_{\phi}S_{\phi}^{{\ell m}} &=&
	\sin\theta\left(\partial^2_{\theta\phi}
	Y^{{\ell m}}-\cot\theta\partial_{\phi}Y^{{\ell m}}\right)\ ,\\
\nabla_{\theta}S_{\phi}^{{\ell m}} &=& 
	\sin\theta \partial^2_{\theta}Y^{{\ell m}}\ ,\\ 
\nabla_{\phi}S_{\theta}^{{\ell m}} &=&
	-\frac{1}{\sin\theta}\partial^2_{\phi}
	Y^{{\ell m}}-\cos\theta\partial_{\theta}Y^{{\ell m}}\ .
\end{eqnarray}
for the vector and tensor odd-parity part. 

	We also recall that the orthogonality relations between scalar,
vector and tensor harmonics imply
\begin{eqnarray}
\int d\Omega Y^*_{{\ell m}}Y_{\ell'm'}& =&\delta_{\ell\ell'}\delta_{m m'}\ ,\\
\int d\Omega \gamma^{cd}(\partial_c Y^*_{\ell m})
	(\partial_d Y_{\ell' m'})&=&\Lambda \delta_{\ell \ell'}
	\delta_{m m'}\ ,\\
\int d\Omega Z^{*{\ell m}}_{cd}Z^{cd}_{{\ell m}}&=&
	\frac{\Lambda(\Lambda-2)}{2}\delta_{\ell \ell'}\delta_{m m'}\ ,
\end{eqnarray}
while the orthogonality relations between the odd-parity axial vectors
$S_{c}$ and tensor $\nabla_{(c} S_{d)}$ follow straightforwardly from the
above equations, giving
\begin{eqnarray}
\int d\Omega \gamma^{cd} S_c^{*\ell m}S_{d}^{\ell'm'}&=&
	\Lambda\delta_{\ell\ell'}\delta_{mm'}\ ,\\
\int d\Omega \nabla_{(c}S^{*\ell m}_{d)}\nabla^{(c}S^{d)}_{\ell' m'}&=&
	 2\Lambda(\Lambda-1)\delta_{\ell\ell'}\delta_{mm'}\ .
\end{eqnarray} 
%

\section{A useful example: point-particles}
\label{sec:particle}

        In the spirit of providing a useful example of a matter-source
generating odd and even-parity perturbations of a Schwarzschild
spacetime, we now a particular problem that has been carefully studied
over the years: that of a point-particle orbiting around a black
hole. Since this problem has been solved in many different cases and
using very different approaches, it can serve as a useful test for codes
handling more complex matter-sources. Because of this, we report below
the relevant components of the odd and even-parity decomposition of the
particle's energy-momentum tensor.

	For a particle of mass $\mu\ll M$, specific energy $E$ and
angular momentum $L$, the equations of motion in the equatorial plane
($\theta=\pi/2$) for the coordinates $R(t)$ and $\Phi(t)$ are
\begin{eqnarray}
{\dot R}^2 &=& \left(1-\frac{2M}{R}\right)^2
	\left[1-\frac{1}{E^2}\left(1-\frac{2M}{R}\right)
	\left(1+\frac{L^2}{R^2}\right)\right]\ ,
\\
{\dot \Phi} &=& \left(1-\frac{2M}{R}\right)\frac{L}{R^2 E}\ ,
\end{eqnarray}
where ${\dot R} \equiv dR/dt$, ${\dot \Phi} \equiv d\Phi/dt$ and where
the energy-momentum tensor reads
\begin{eqnarray}
\fl t_{\mu\nu}&=&\frac{\mu}{r^2E}\left(1-\frac{2M}{r}\right)
	\delta(r-R(t))\delta(\phi-\Phi(t))
	         \delta\left(\theta-\frac{\pi}{2}\right)\times
\\
\fl &\times&
\left(\begin{array}{cccc} E^2 &
	E^2\left(1-2M/r\right)^{-2}{\dot R} &
	 0   &
	-LE \\ 
	\\
	{\rm sym} & E^2\left(1-2M/r\right)^{-4}{\dot R}^2 &
	0   & 
	LE \left(1-2M/r\right)^{-2}{\dot R}\\
	\\
	{\rm sym} & {\rm sym} & 0  &    0 \\
	\\
	{\rm sym} & {\rm sym} & {\rm sym} & L^2 
\end{array}
\right)\ .
\end{eqnarray}
As a result, the odd-parity components of the stress energy tensor
are\footnote{Note that the spherical harmonics and their derivatives are
all evaluated at $\phi=\Phi(t)$ and $\theta=\pi/2$.}
\begin{eqnarray}
L^{\ell m}    &=& -\frac{{\rm i} m}{\Lambda(\Lambda-1)}
	\frac{\mu L^2}{r^2 E}\left(1-\frac{2M}{r}\right)
	\;\delta(r-R(t))\de_{\theta}Y^*_{\ell m}\ ,
\\
L_0^{{\ell m}}&=&-\frac{1}{\Lambda}\frac{\mu L}{r^2}
	\left(1-\frac{2M}{r}\right)\delta(r-R(t))
	\de_{\theta}Y^{*}_{{\ell m}}\ ,
\\
L_1^{{\ell m}}&=& \frac{1}{\Lambda}\frac{\mu L}{r^2}
	\left(1-\frac{2M}{r}\right)^{-1}{\dot R}
	\;\delta(r-R(t))\de_{\theta}Y^{*}_{{\ell m}}\ ,
\end{eqnarray}
while the even-parity components assume the form
\begin{eqnarray}
T_{00}^{{\ell m}}&=&\mu E \frac{r-2M}{r^3}\delta(r-R(t))Y^*_{{\ell
        m}}\ ,\\
T_{11}^{{\ell m}}&=& \frac{\mu
        Er}{(r-2M)^3}\,({\dot R})^2\,\delta(r-R(t))Y^*_{{\ell
        m}}\ ,\\
T_{01}^{{\ell m}}&=&
        -\frac{\mu}{r(r-2M)}\,{\dot R}\,\delta(r-R(t))Y^*_{{\ell m}}\ ,\\
T_0^{{\ell m}} &=& \frac{\mathrm{i}m\mu
        L}{\Lambda}\,\frac{r-2M}{r^3}\,\delta(r-R(t))Y^*_{{\ell m}}\ ,\\
T_1^{{\ell m}} &=& -\mathrm{i} m\frac{\mu L}{\Lambda
        r(r-2M)}\,{\dot R}\,\delta(r-R(t))Y^*_{{\ell m}}\ ,\\
T_2^{{\ell m}} &=&
        \frac{\Lambda-2m^2}{\Lambda(\Lambda-2)}\,\frac{\mu L^2 (r-2M)}{E
        r^3}\,\delta(r-R(t))Y^*_{{\ell m}}\ ,\\
T_3^{{\ell m}} &=& \frac{\mu
        L^2(r-2M)}{2Er^5}\,\delta(r-R(t))Y^*_{{\ell m}}\ .
\end{eqnarray}

\section{Variational principle for odd-parity perturbations}
\label{appendix_d}

	In Sect.~\ref{opp} we have discussed that it is possible to
choose two different representations for the odd-parity perturbations,
namely expressions (\ref{CPM_n}) for the CPM convention and expression
(\ref{mfop}) for the RW convention. An intimate connection links the two
master functions and this can be appreciated by using the variational
formalism employed by Moncrief in~\cite{moncrief74}. To highlight this,
let us recall that two perturbation functions $k_1, k_2$ can be
introduced for the odd-parity perturbations
\begin{eqnarray}
\label{k1:monc}
\fl \hskip 1.5 cm 
k_1&\equiv h_1^{(\rm o)}+\frac{1}{2}\left(\de_rh_2-
	\frac{2}{r}h_2\right)\ , \hskip 2.5cm
\label{k2:monc}
	k_2&\equiv h_2 = -2 h\ ,
\end{eqnarray}
where $k_1$ is gauge-invariant (cf equation~(\ref{def:kA})) but $k_2$ is not. 
While this may seem just another possible choice for the odd-parity 
perturbations, it points out that it is now possible to introduce the 
quantities $\pi_1$ and $\pi_2$ as conjugate momenta of $k_1$ and $k_2$ 
in the Hamiltonian
\begin{eqnarray}
\label{hamiltonian}
\fl H_{\rm T} \equiv \int dr{\cal H} &=&\frac{1}{\Lambda}
	\int dr\left[\frac{1}{2}\pi_1^2+\frac{2r(r-2M)}{(\Lambda - 2)}
	\left(\pi_2-\frac{1}{2}
	\de_r\pi_1-\frac{1}{r}\pi_1\right)^2\right]
\nonumber\\
\fl && \hskip 0.0 cm +\frac{1}{2}\Lambda\int dr\left[\frac{\Lambda - 2}{r^2}
	\left(1-\frac{2M}{r}\right)k_1^2\right]-2\int dr h_0^{(\rm o)}\pi_2\;.
\end{eqnarray}
In addition to the constraint equation $\pi_2=0$ (which must be enforced
at initial time and is then conserved since $k_2$ is cyclic), the other
Hamilton equations are
\begin{eqnarray}
\label{eqs1}
\fl 	
\hskip 1.5 cm \de_tk_1 = \frac{\delta H_{\rm T}}{\delta \pi_1}\ ,
\hskip 1.5 cm 
	\de_tk_2 = \frac{\delta H_{\rm T}}{\delta \pi_2}\ ,
\hskip 1.5 cm 
	\de_t\pi_1 =-\frac{\delta H_{\rm T}}{\delta k_1}\ ,
\end{eqnarray}
where $\delta(\;\,)/\delta(\;\,)$ refers to a functional differentiation.
The first and third of eqs.~(\ref{eqs1}) have explicit expressions
\begin{eqnarray}
\label{eq:dk1dt}
\fl \de_tk_1&=&\frac{1}{\Lambda(\Lambda - 2)}
	\left[(\Lambda - 2)\pi_1-\left(1-\frac{2M}{r}\right)
	\partial^2_{r}(r^2\pi_1)
	+\frac{2}{r}\left(1-\frac{3M}{r}\right)\de_r(r^2\pi_1)\right]\ ,
\\
\label{eq:dpi1dt}
\fl \de_t\pi_1 &=& -\frac{\Lambda(\Lambda - 2)}{r^2}
	\left(1-\frac{2M}{r}\right)k_1\ .
\end{eqnarray}
and the Regge-Wheeler equation in the RWM convention can be obtained by
time-differentiating equation~(\ref{eq:dk1dt}), replacing $\de_t\pi_1$ through
equation~(\ref{eq:dpi1dt}) and introducing the quantity
\begin{equation}
Q^{(\rm o)}=\frac{k_1}{r}\left(1-\frac{2M}{r}\right)\ ,
\end{equation}
to obtain
\begin{equation}
\label{RWM_eq}
\de^2_t Q^{(\rm o)}-\de^2_{r_*}Q^{(\rm o)}+
	\left(1-\frac{2M}{r}\right)
	\left(\frac{\Lambda}{r^2}-
	\frac{6M}{r^3}\right)Q^{(\rm o)}=0\ .
\end{equation}
Equation~(\ref{RWM_eq}) coincides with equation~(4.20) 
of~\cite{moncrief74} with $Q = Q^{(\rm o)}$.

	Remarkably, the Regge-Wheeler equation in the CPM convention can
be obtained also through the Hamilton equations in terms of the
conjugate moment to $k_1$, i.e.  $r\pi_1$. More specifically, it is
sufficient to differentiate in time equation~(\ref{eq:dpi1dt}) and to replace
$\de_t k_1$ in equation~(\ref{eq:dk1dt}) to obtain
\begin{equation}
\label{CPM_equation}
\de^2_t (r\pi_1)-\de^2_{r_*}(r\pi_1)+\left(1-\frac{2M}{r}\right)
	\left(\frac{\Lambda}{r^2}-\frac{6M}{r^3}\right)r\pi_1=0\ ,
\end{equation}
Equation~(\ref{RWM_eq}) coincides with equation~(II-15) of~\cite{CPM78}
with
\begin{equation}
\label{eq:CPM_II-14}
r \pi_1 = \frac{\widetilde{\psi}}{\Lambda} \equiv
	r \left[\de_t h_1^{(\rm o)}-
	r^2\de_r \left(\frac{h_0^{(\rm o)}}{r^2}\right)\right]\ .
\end{equation}
Note also that the second of Hamilton equations (\ref{eqs1}) can be used
to derive
\begin{equation}
\label{eq:III-20_CPM_v1}
\de_t h_2=-\frac{2}{\Lambda(\Lambda - 2)}
	\left(1-\frac{2M}{r}\right)
	\de_r\left(r^2\pi_1\right)-2h_0^{(\rm o)}\ ,
\end{equation}
which coincides with equation~(III-20) of Cunningham {\it et al.}~\cite{CPM78}
and has been used to obtain the asymptotic relation between $\Psi^{(\rm
o)}$ and the gravitational wave amplitude (cf equation~(\ref{eq:III-20})).

	Finally, we underline that the relation (\ref{Q_vs_Psi}) between
the two master functions $\Psi^{({\rm o})}$ and $Q^{({\rm o})}$ can be
obtained in coordinate independent form also in the formalism introduced
by Mart\'in-Garc\'ia and Gundlach~\cite{martingarcia99}. More
specifically, exploiting the odd-parity perturbations equations
(\ref{odd_constr_eq1}) as
\begin{eqnarray}
\fl g{\!\!\!\!\; \raisebox{-0.1ex}{$^{^0}$}}^{CD}\nabla_{\!D} 
	\left[4r(\nabla_{\![A} r) k_{C]} + 
	2 r^2 \nabla_{\![C} k_{A]}\right]n^A
	- (\Lambda-2) k_{A}n^A = -16\pi L^{\ell m}_{A}n^A
\end{eqnarray}
and the gauge-invariant and coordinate independent definitions of
$\Psi^{(\rm o)}$ and $Q^{(\rm o)}$ (i.e. equations~(\ref{phi:GM}) 
and (\ref{mfop}) of the main text), straightforward calculations then 
yield
\begin{equation}
\label{relation_mgg}
\dot{\Psi}^{(\rm o)}=-\frac{\,\,\,Q^{(\rm o)}}{r'}
  	+\frac{16\pi}{\Lambda-2} r n^A L_A^{\ell m}
\end{equation} 
where $\dot{\Psi}^{({\rm o})} \equiv u^A\nabla_{\!A}\Psi^{({\rm o})}$ and
$r' \equiv n^A \nabla_{\!A} r $ are `frame derivatives' in terms of the
non-coordinate basis vectors $(u^A,n^A)$ of ${\sf M}^2$, such that
$-u_Au^A=n_An^A=1$, $g{\!\!\!\!\; \raisebox{-0.1ex}
{$^{^0}$}}_{AB}=-u_Au_B+n_An_B$ and $\epsilon_{AB}=n_Au_B-u_An_B$. 
In the case of Schwarzschild coordinates, $n^A = (0,e^{-b})$, 
$u^A = (e^b,0)$ and equation~(\ref{relation_mgg}) coincides with 
equation~(\ref{Q_vs_Psi}).

\section{Angular pattern functions}

	Finally, this section is devoted to a list of the expressions of
the angular functions $W^{{\ell m}}$ and $X^{{\ell m}}$ for the first
values of $\ell$ and $m$. While these expressions can be easily
reproduced with straightforward algebra, they are tedious to derive and
hard to find in the literature.

\bigskip 
$\bullet \quad \ell=2$ 
\begin{eqnarray} 
\fl W^{20} &=&
	\sqrt{\frac{45}{4\pi}}\sin^2\theta\,,\qquad\qquad\qquad \fl
	W^{21} = e^{\mathrm{i}\phi}\sqrt{\frac{15}{8\pi}}\sin
	2\theta\,,\qquad\qquad\qquad \fl W^{22} =
	e^{\mathrm{i}2\phi}\sqrt{\frac{15}{32\pi}}\left(3+\cos
	2\theta\right)\,\\ \fl X^{20} &=&
	0\,,\qquad\qquad\qquad\quad\qquad\quad \fl X^{21} =
	\mathrm{i}e^{\mathrm{i}\phi}\sqrt{\frac{15}{2\pi}}\sin^2\theta\,,
	\qquad\qquad\quad\quad\;\,
	\fl X^{22} = \mathrm{i}
	e^{\mathrm{i}2\phi}\sqrt{\frac{15}{8\pi}}\sin 2\theta\ .
\end{eqnarray} 

\bigskip 
$\bullet \quad \ell=3$ 
\begin{eqnarray} 
\fl W^{30} &=&
	\frac{15}{2}\sqrt{\frac{7}{\pi}}\cos\theta\sin^2\theta\ ,
	\qquad\qquad\quad
	W^{31} = e^{\mathrm{i}\phi}\frac{5}{8}
	\sqrt{\frac{21}{\pi}}\left(1+3\cos
	2\theta\right)\sin\theta,\\ \fl W^{32} &=&
	e^{\mathrm{i}2\phi}\sqrt{\frac{105}{32\pi}}\left(1+3\cos
	2\theta\right)\cos\theta\ ,
	\quad
	W^{33} = -e^{\mathrm{i}3\phi}\sqrt{\frac{315}{64\pi}}\left(3+\cos
	2\theta\right)\sin\theta \ ,\\ 
	\fl X^{31} &=& \mathrm{i}
	e^{\mathrm{i}\phi}\sqrt{\frac{525}{4\pi}}\cos\theta\sin^2\theta\ ,
	\qquad\qquad
	X^{32} = \mathrm{i}e^{\mathrm{i}2\phi}
	\sqrt{\frac{105}{2\pi}}\sin\theta\cos
	2\theta\ ,  \\ 
	\fl X^{33} &=& -\mathrm{i}e^{\mathrm{i}3\phi}
	\frac{3}{2}\sqrt{\frac{35}{\pi}}\cos\theta\sin^2\theta\ .
\end{eqnarray} 

\bigskip 
$\bullet \quad \ell=4$ 
\begin{eqnarray} 
	\fl W^{40} =
	\frac{45}{8\sqrt\pi}\left(5+7\cos 2\theta\right)\sin^2\theta\ ,
	\hskip 3.0cm
	\fl W^{41} =
	e^{\mathrm{i}\phi}\frac{9}{8}\sqrt{\frac{5}{\pi}}\left(7\cos
	2\theta -1\right)\sin 2\theta\ ,\\ 
	\fl W^{42} =
	e^{\mathrm{i}2\phi}\frac{9}{16}\sqrt{\frac{5}{2\pi}}\left(5+4\cos
	2\theta+7\cos 4\theta\right)\ ,
	\hskip 3.0cm
	\fl W^{43} =
	-\mathrm{i}e^{\mathrm{i}3\phi}\frac{9}{2}
	\sqrt{\frac{35}{\pi}}\sin\theta \cos^3\theta\ ,
	\nonumber \\  \\
	\fl W^{44} =
	e^{\mathrm{i}4\phi}\frac{9}{8}\sqrt{\frac{35}{2\pi}}\left(3+\cos
	2\theta\right)\sin^2\theta\ ,
	\hskip 3.0cm
	\fl X^{41} =
	\mathrm{i}e^{\mathrm{i}\phi}\frac{9}{8}\sqrt{\frac{5}{\pi}}\left(7\cos
	2\theta+5\right)\sin^2\theta\ ,\\ 
	\fl X^{42} =
	\mathrm{i}e^{\mathrm{i}2\phi}\frac{9}{8}
	\sqrt{\frac{5}{2\pi}}\left(7\cos
	2\theta-3\right)\sin 2\theta\ ,
	\hskip 3.0cm
	\fl X^{43} =
	-\mathrm{i}e^{\mathrm{i}3\phi}
	\frac{9}{8}\sqrt{\frac{35}{\pi}}\left(1+3\cos
	2\theta\right)\sin^2\theta\ , 
	\nonumber \\  \\
	\fl X^{44} =
	\mathrm{i}e^{\mathrm{i}4\phi}\frac{9}{2}
	\sqrt{\frac{35}{2\pi}}\cos\theta\sin^2\theta\ .
\end{eqnarray} 

\bigskip 
$\bullet \quad \ell=5$ 
 \begin{eqnarray} \fl W^{50} &=&
 	\frac{105}{16}\sqrt{\frac{11}{\pi}}\left(5\cos\theta+3\cos
 	3\theta\right)\sin^2\theta\ ,\\ \fl W^{51} &=& e^{{\rm
 	i}\phi}\frac{7}{32}\sqrt{\frac{165}{2\pi}}\left(5+12\cos
 	2\theta+15\cos 4\theta\right)\sin\theta\ , \\ \fl W^{52} &=&
 	e^{{\rm i}2\phi}\sqrt{\frac{1155}{512\pi}}\left(13-12\cos
 	2\theta+15\cos 4\theta\right)\cos\theta\ , \\ \fl W^{53} &=&
 	-e^{{\rm
 	i}3\phi}\frac{3}{64}\sqrt{\frac{385}{\pi}}\left(21+28\cos
 	2\theta+15\cos 4\theta\right)\sin\theta\ , \\ \fl W^{54} &=&
 	e^{{\rm i}4\phi}3\sqrt{\frac{385}{128\pi}}\left(7+5\cos
 	2\theta\right)\cos\theta\sin^2\theta\ , \\ \fl W^{55} &=&
 	-e^{{\rm i}5\phi}\frac{15}{16}\sqrt{\frac{77}{\pi}}\left(3+\cos
 	2\theta\right)\sin^3\theta\ ,\\ \fl X^{50} &=& 0 \ ,\\ \fl X^{51}
 	&=& {\rm i}e^{{\rm
 	i}\phi}\frac{7}{4}\sqrt{\frac{165}{2\pi}}\left(1+3\cos
 	2\theta\right)\sin^2\theta\ ,\\ \fl X^{52} &=& {\rm i} e^{{\rm
 	i}2\phi}\sqrt{\frac{1155}{32\pi}}\left(\cos 2\theta+3\cos
 	4\theta\right)\sin\theta\ ,\\ \fl X^{53} &=& -{\rm i}e^{{\rm
 	i}3\phi}3\sqrt{\frac{385}{64\pi}}\left(9\cos
 	2\theta-1\right)\cos\theta\sin^2\theta\ ,\\ \fl X^{54} &=& {\rm
 	i}e^{{\rm i}4\phi}3\sqrt{\frac{385}{8\pi}}\left(1+2\cos
 	2\theta\right)\sin^3\theta\ ,\\ \fl X^{55} &=& -{\rm i} e^{{\rm
 	i}5\phi}15\sqrt{\frac{77}{16\pi}}\cos\theta\sin^4\theta\ .
 	\end{eqnarray}

\section*{References}

\end{document}